\documentclass[reprint,
prl,amsmath,amssymb,aps,superscriptaddress]{revtex4-2}

\usepackage{graphicx}
\usepackage{dcolumn}
\usepackage{bm}

\usepackage{circuitikz}
\usepackage{natbib}
\usepackage{simpler-wick}[sep = 5pt, offset = 5em]   
\usepackage[caption=false]{subfig}
\usepackage{siunitx}
\begin{document}
\title{\Large Light-Matter Interaction in dispersive Superconducting Circuit QED}
\author{Harsh Arora}
\affiliation{Center for Nano Science and Engineering, Indian Institute of Science, Bengaluru, India - 560012}
\author{Jay Deshmukh}
\affiliation{Department of Condensed Matter Physics, Tata Institute for Fundamenetal Research, Mumbai, India - 400005}
\author{Ansh Das}
\affiliation{Birla Institute of Technology and Science, Pilani, India - 333031}
\author{R. Vijay}
\affiliation{Department of Condensed Matter Physics, Tata Institute for Fundamenetal Research, Mumbai, India - 400005}
\author{Baladitya Suri}
\affiliation{Instrumentation and Applied Physics, Indian Institute of Science, Bengaluru, India - 560012}
\begin{abstract}
It is well known that superconducting waveguides strongly attenuate the propagation of electromagnetic waves with frequencies beyond the superconducting gap. In circuit QED, the interaction between non-linear charge qubits and superconducting resonators invariably involves the qubit coupling to a large set of resonator modes. So far, strong dispersion effects near and beyond the superconducting-gap have been ignored in quantization models. Rather, it is assumed that the superconducting resonator behaves ideally across the large frequency intervals. We present a quantization approach which includes the superconducting frequency-dependent surface impedance and demonstrate that superconducting dispersion plays a role in determining the effective light-matter interaction cut-off. 
\end{abstract}
\maketitle
In circuit-QED (cQED) systems, consisting of nonlinear qubits coupled to superconducting resonators, it has been observed that considerably de-tuned off-resonant modes of the resonator can have a non-trivial impact on the qubit computational sub-space\cite{houckdispersive,Houckoffresonant}. Since the capacitive coupling $g_{n}$ between the resonator and qubit scales as square root of the mode frequency\cite{HakanA2}, $g_{n} \sim \sqrt{\omega_{n}}$, the de-tuning of the resonator modes is off-set by the increase in coupling strengths. Theoretically, a natural cut-off occurs because the potential difference across the josephon junction is suppressed at high frequencies for capacitive couplings\cite{Hakan}. This suppression occurs when the impedance of the series capacitance, $C_{s} = \frac{C_{g}C_{J}}{C_{g} + C_{J}}$ formed by junction shunt capacitance $C_{J}$ and the coupling capacitance $C_{g}$ is less than the impedance of the CPW resonator\cite{GS}: 
\begin{eqnarray}
    \frac{1}{\omega C_{s}} \leq 50 \;\text{Ohms}
\end{eqnarray}
leading to $\omega \sim O(10^{3})$ GHz for a fairly standard value of $C_{s} \sim 10^{-14}$ F. However, these estimated values are greatly beyond the superconducting-gap of commonly utilized BCS superconductors, raising concerns about the physical relevance of these high-frequency cut-offs \footnote{Additionally, for high frequencies the dipole approximation breaks down since the wavelength of the CPW mode is comparable to the transmon's spatial extent, resulting in the mode averaging out to zero over the transmon\cite{WallraffMultimode} for $\omega >> 10^{4}$ GHz}.  
\\\\
Further, superconducting circuits exhibit dispersion and dissipation of the electromagnetic field when operated over a wide band of microwave frequencies\cite{kautz1978picosecond,chi1987subpicosecond,chi1988tera,gallagher1987subpicosecond}. As is known from the Meissner effect, the tangential components of the electric and magnetic fields decay exponentially into the superconductor with the scale set by the London penetration depth $\lambda_{L}$. Therefore, a fraction of the electromagnetic energy is stored \& dissipated by the superconducting electrons in a region close to the metal's surface as $\lambda_{L} \sim O(10^{-9})$m for type-I BCS superconductors\cite{gaothesis} such as Aluminum, Niobium \& Tantalum. Additionally, oxides formed on the metal as well as the substrate surface during fabrication hosts several Two-Level System defects\cite{muller2019review,martinis2005decoherence,gaothesis}, which interact with the electric field and lead to further degradation of the circuit's operation. For example, the fidelity of readout operations is reduced due to losses introduced by an excess of high-energy quasi-particles generated by the applied microwave pulses\cite{excessQPmicrowave1,excessQPmicrowave2}. In large scale quantum processors, TLS defects in junction oxides and interfaces lead to a drift in the qubit and resonator frequencies and decay rates over the time-scales of hours\cite{klimov2018fluctuations,resonator1,resonator2,resonator4}, leading to increased overheads\cite{klimov2020snake}. 
\\\\
Interestingly, disordered superconductors\cite{GrAlFilms,GrAlStates,GrAlSuperconductivity} offer a differing narrative. The high surface impedance of Granular Aluminum films enables compact, high-inductance resonators\cite{robdisorder}, which are crucial for Fluxonium qubits\cite{GrAlFluxqubitr}, sensitive Kinetic Inductance detectors\cite{GrAlMKID,GrAlMKID2}, and hybrid superconductor-semiconductor quantum devices\cite{GrAlNovel1,GrAlNovel2}. Furthermore, introducing dispersive elements in superconducting resonators facilitates the engineering of meta-materials\cite{metamaterial}  with several applications, such as photonic crystals\cite{photoniccrystal} \& Josephson traveling wave parametric amplifiers\cite{grimsmo2017squeezing}.  Therefore, the dispersive nature of superconducting materials is important for both - loss mitigation and the development of new devices. 
\\\\
In this letter we revisit the canonical quantization procedure for cQED systems. We present an analytical framework for capturing the effects of the surface impedance of superconducting resonators. The quasi-TEM nature of the Coplanar waveguide resonator's field distribution implies that a component of the field Poynting vector is directed into the superconducting metal\cite{gao2008physics}. Therefore,  modes above the superconducting-gap have a finite lifetime since they can break the Cooper pairs into quasi-particles, effectively dissipating energy via the creation of phonons\footnote{The quasi-particles have a finite lifetime and recombine into Cooper pairs via emission of phonons\cite{wang2014measurement,Recomb1,Recomb2,Recomb3}}. To capture these effects, we extend the Mattis-Bardeen conductivity\cite{MB} to the complex plane and demonstrate that the mode's finite lifetime effectively acts to decouple the qubit and the resonator. Our work also holds relevance for other cQED applications, such as pulse propagation\cite{varvelis2024photonic} and quantum state transfer\cite{QST}, which involve a large frequency range of the resonator modes.  
\\\\
Consider the generalized distributed element model\cite{HeinrichCPW} for superconducting resonators, Fig. 1.
\begin{figure}[h]
\begin{circuitikz}
\draw (-1,2)node[below]{$x-\Delta x$} to [generic, l = $Z(\omega)$, *-*](2,2)node[above]{$x$};
\draw (2,2) to [generic, l= $Z(\omega)$,-*](5,2)node[below]{$x +\Delta x$};
\draw (2,2) to [C,l = $c(x)\Delta x$](2,0.3)node[ground]{};
\end{circuitikz}
\caption{\label{CPW} Distributed element model for resonators incorporating the frequency-dependent elements: For Coplanar waveguides (CPW), $g$ is determined by width of the center conductor s, separation between the ground planes and thickness of the metal film.} 
\end{figure}
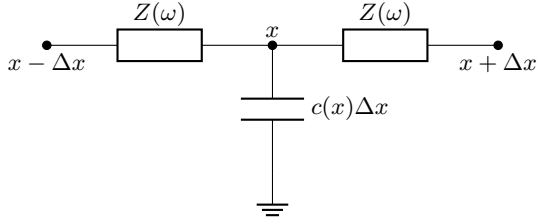
We express the total impedance per unit length as $Z(\omega) = i\omega\ell_{m} + gZ_{s}(\omega)$, where $Z_{s}(\omega)$ is the superconductor's complex contribution to the impedance per unit and $\ell_{m}$ is the magnetic inductance per unit length. The factor $g$ is a geometric factor which takes into account the resonator geometry\cite{gao2008physics}. In general, the surface impedance $Z_{s}(\omega) = R_{s}(\omega) + iX_{s}(\omega)$ is complex: $R_{s}(\omega)$ represents the surface resistance and $X_{s}(\omega)$ represents the surface reactance due to the Kinetic Energy stored by the Cooper pairs.  Assigning node fluxes $\phi(x_{j},t) = \int d\tau V(x_{j},\tau)$ and applying Kirchoff's laws at the nodes, we get the following wave-equation for the node flux in the frequency domain:
\begin{align}
     \frac{\partial^{2} \tilde{\phi}(x,\omega)}{\partial x^{2}} + \omega^{2}c(x)\ell_{m}\epsilon(\omega)\tilde{\phi}(x,\omega) = 0
\end{align}
where we define $\tilde{\phi}(x,\omega) = \int_{-\infty}^{\infty} d\omega e^{i\omega t}\phi(x,t)$. The above equation describes wave-propagation in a media with a complex refractive index $\epsilon(\omega)$ given as: 
\begin{eqnarray}
   \epsilon(\omega) \equiv  1 + \frac{g}{i\omega \ell_{m}}Z_{s}(\omega) = 1 + \frac{gX_{s}(\omega)}{\omega\ell_{m}} - i\frac{gR_{s}(\omega)}{\omega\ell_{m}}
\end{eqnarray}
Equation $(1)$ is non-hermitian for frequency intervals where surface resistance cannot be ignored. Since the solutions are exponentially damped in space, a straightforward quantization procedure in terms of the eigenmodes is not applicable. We resort to a non-perturbative Open Quantum System approach to account for the losses by considering a reservoir coupled to the fluxes $\phi(x,t)$. We find that the Hamiltonian for a finite length CPW interacting with a total of $M$ capacitively coupled charge qubits is given by\cite{supp} $H = \sum_{j=1}^{j=M}\frac{1}{2C_{J_{j}}}Q_{J_{j}}^{2} - E_{J_{j}}\cos\left(2\pi\frac{\Phi_{J_{j}}}{\phi_{o}}\right) + \mathcal{H}_{int} + \int dx \mathcal{H}_{TL,R}  $ with the Hamiltonian density $\mathcal{H}_{TL,R}$ given as $\mathcal{H}_{TL,R}=$: 
\begin{eqnarray}
&&\frac{1}{2c(x)}\left(\rho(x) + \sqrt{\frac{2}{\pi}}\int_{0}^{\infty}d\omega \sqrt{\frac{gR_{s}(\omega)c(x)}{\ell_{m}}}\psi_{\omega}(x)\right)^{2} \nonumber \\&& + \frac{1}{2\ell_{m}} \left(\frac{\partial \phi}{\partial x}\right)^{2}  +  \frac{1}{2}\int_{0}^{\infty} d\omega \rho_{\omega}^{2}(x) + \omega^{2}\psi_{\omega}^{2}(x) 
 \end{eqnarray}
where $\rho_{\omega}(x)$ \& $\psi_{\omega}(x)$ denote the reservoir momenta and displacement fields. In the classical limit, the above Hamiltonian recovers wave-equation $(2)$\cite{supp}. Due to the the Kramer-Kronig relations necessarily satisfied by $\epsilon(\omega)$, the interaction in $(4)$ leads to pure dispersion for the Resonator modes whenever $R_{s}(\omega) = 0$. The Qubit-Resonator interaction takes the form $H_{int} = \sum_{j=1}^{j=M}\gamma_{j}Q_{J_{j}}\frac{\rho(x,t)\delta(x-x_{j})}{c(x)}
$, being proportional to the product of charges on the Qubit and the Resonator. The position of the qubit and their coupling strengths are encoded in the capacitance per unit length $c(x) = c + \sum_{j=1}^{j=M}C_{s_{j}}\delta(x-x_{j})$, where $C_{s_{j}}$ is the total series capacitance to ground at the qubit location. Therefore, the Hamiltonian $\mathcal{H}_{TL,R}$ is a generalization of the Current-conserving (CC) Hamiltonian considered previously\cite{HakanA2}.
\\\\
\textit{Canonical quantization:} The correct basis for considering the cQED light-matter interaction is given by the solutions to Hamilton equations of motion obtained from the resonator-reservior Hamiltonian $\mathcal{H}_{TL,R}$. We find that these solutions are the eigenmodes of the hermitian wave-equation for frequencies in which $R_{s} =0$:   
\begin{align}
    \omega^{\prime 2}\ell_{m}c(x)Re\{\epsilon(\omega)\}\tilde{\phi}(x,\omega^{\prime}) + \frac{\partial^{2}\tilde{\phi}(x,\omega^{\prime})}{\partial x^{2}} = 0
\end{align}
Otherwise, for frequency intervals with non-zero $R_{s}(\omega)$ the node fluxes satisfy a non-homogeneous equation:
\begin{align}
    &\omega^{ 2}\ell_{m}c(x)\epsilon(\omega)\tilde{\phi}(x,\omega) + \frac{\partial^{2}\tilde{\phi}(x,\omega)}{\partial x^{2}} = \mathbf{j}_{\text{noise}}(x,\omega)
\end{align}
where $\mathbf{j}_{noise} = -i\sqrt{g\ell_{m}R_{s}(\omega)c(x)}\omega\mathbf{X}_{\omega}(x)$, with $\mathbf{X}_{\omega}(x)$ representing the zero-point fluctuations (ZPF) of the reservoir\cite{supp}. Then, the modification to the flux and charge density due to the dispersive nature of the resonator is captured by Eqs. $(5)$ \& $(6)$. For lossless propagation, the normal modes are the hermitian eigenmodes of Eq. $(2)$. Eq. $(6)$ reveals that the fluxes are  driven by the reservoir ZPF's and can be calculated  using the Green's function  $G(x,x^{\prime};\omega)$ of the non-hermitian wave-equation $(2)$. The Green's function, arising in the study of local perturbations to the resonator electromagnetic field, completely captures the dynamical effect of the qubit on the resonator modes, and vice versa\cite{NMHakan}. As is well known, the frequency response of a damped harmonic oscillator is maximum at it's bare frequency, and diminishes with increasing damping. Similarly, the Green's function of the lossy wave-equation $(2)$ effectively acts to decouple the qubit \& resonator with increasing losses which arise due to the surface resistance, $R_{s}(\omega)$. We calculate $G(x,x^{\prime};\omega)$ using the bi-orthogonal eigenmodes $\{\Psi_{n}, \tilde{\Phi}_{n}\}$ of Eq. $(2)$:
\begin{eqnarray}
    G(x,x^{\prime};\omega) = \sum_{n}\frac{\Psi_{n}(x)\tilde{\Phi}^{\star}_{n}(x^{\prime})}{\omega^{2} - \omega_{n}^{2}(\omega)}
\end{eqnarray} 
where these bi-orthgonal eigenmodes satisfy the following generalized eigenvalue equation\cite{morse1946methods}: 
\begin{eqnarray}
 \left(\frac{1}{\ell_{m}c(x)}\partial_{x}^{2}   -\frac{ig\omega Z_{s}(\omega)}{\ell_{m}}\right)\Psi_{n} &&= -\omega_{n}^{2}(\omega)\Psi_{n} \\
\left(\frac{1}{\ell_{m}c(x)}\partial_{x}^{2} + \frac{ig\omega Z_{s}^{\star}(\omega)}{\ell_{m}}\right)\tilde{\Phi}_{m} &&= -\omega_{m}^{\prime 2}(\omega)\tilde{\Phi}_{m}
\end{eqnarray}
Generally, Eqs. $(8)$ \& $(9)$ are solved with radiative boundary conditions, signifying loss through the ends of the resonator. However, above the superconducting-gap the quasi-particle loss greatly exceeds other loss sources. Hence, we consider open boundary conditions\footnote{The generalization to periodic boundary conditions, arising in the context of ring-resonator based couplers is straightforward\cite{ringresonator}} $\frac{\partial \Psi_{n}}{\partial x}\big|_{0, L} = 0$. Re-arranging the eigenmode equation as $\frac{\partial^{2}\Psi_{n}}{\partial x^{2}} = \ell_{m}c(x)\left(-\omega_{n}^{2}(\omega) + \frac{ig\omega Z_{s}(\omega)}{\ell_{m}}\right)\Psi_{n}(x)$; we find that the generalized eigenvalues are fully determined by the geometry of the resonator, encoded entirely into the boundary conditions and the capacitance per unit length $c(x)$. Defining the wave-vector as $\frac{k_{n}^{2}}{\ell_{m}c} = \omega_{n}^{2}(\omega) - \frac{ig\omega Z_{s}(\omega)}{\ell_{m}}$ we find that the eigenmodes are given by solutions of $\frac{\partial^{2}\Psi_{n}}{\partial x^{2}} = -k_{n}^{2}\Psi_{n}(x)$, with $k_{n}$ determined entirely by the boundary conditions and the CC conditions imposed at the transmon positions\cite{supp}. Once the $k_{n}$ are determined the generalized eigen-frequencies are easily found; however, the quasi-bound solutions $\omega = \omega_{n}(\omega)$ which are the poles of the Green's function determine the loss rates and the eigen-frequencies of the lossy modes. Therefore, we are interested in the fixed points of the dispersion relation: 
\begin{align}
    \omega_{n}^{2}= \frac{1}{\ell_{m}c}\left(k_{n}^{2} + ig\omega_{n} c Z_{s}(\omega_{n})\right)
\end{align}
In intervals where $R_{s}(\omega)$ cannot be ignored the fixed points are complex. Then, the overall solution for the flux $\phi(x,t)$ and charge density $\rho(x,t)$ is given by the Fourier transform of the solution to Eqs. $(5)$ and $(6)$, each holding in different frequency intervals. At zero temperature, we have the following solutions, general to all geometries: 
\begin{widetext}
    \begin{eqnarray}
    \hat{\phi}(x) &=& \sum_{\omega_{n} < \frac{2\Delta}{\hbar}}\sqrt{\frac{\hbar\ell_{m}}{2\omega_{n}}}\sqrt{\epsilon(\omega_{n})}\tilde{\Psi}_{n}(x,\omega_{n})\hat{a}_{n}  -i\sqrt{\frac{\hbar}{\pi}}\int_{\frac{2\Delta}{\hbar}}^{\infty}d\omega \int_{0}^{L} dx \sqrt{\omega}G(x,x^{\prime};\omega)\sqrt{g\ell_{m}R_{s}(\omega)c(x^{\prime})}\hat{f}(x^{\prime},\omega)  + \text{H.C.} \\
        \hat{\rho}(x) &=& -i\sum_{\omega_{n} < \frac{2\Delta}{\hbar}}\sqrt{\frac{\hbar\ell_{m}\omega_{n}}{2}}\sqrt{\epsilon(\omega_{n})}\tilde{\Psi}_{n}(x,\omega_{n})\hat{a}_{n} + c(x)\int_{\frac{2\Delta}{\hbar}}^{\infty} d\omega \int_{0}^{L}dx^{\prime}\omega^{3/2}G(x,x^{\prime};\omega)\epsilon(\omega)\sqrt{g\ell_{m}R_{s}(\omega)c(x^{\prime})}\hat{f}(x^{\prime},\omega) \nonumber \\ && + \sqrt{\frac{\hbar}{\pi}}\int_{\frac{2\Delta}{\hbar}}^{\infty}d\omega\sqrt{\frac{2 gR_{s}(\omega)c(x)}{\pi\omega\ell_{m}}}\hat{f}(x,\omega)  + \text{H.C.} 
         \end{eqnarray}
     \end{widetext}
The operators $\hat{a}_{n}$, $\hat{a}_{n}^{\dagger}$ correspond to the the annihilation and creation operators for lossless modes below the superconducting gap and satisfy $[\hat{a}_{n},\hat{a}^{\dagger}_{n}]=1$. Above $2\Delta$, a continuum of modes description emerges as traveling wave solutions do not satisfy the Heisenberg equations of motion for $R_{s}(\omega) \neq 0$. The operators $\hat{f}(x,\omega)$, $\hat{f}^{\dagger}(x,\omega)$ correspond to the annihilation and creation operators for the reservoir modes and satisfy: $[\hat{f}(x,\omega),\hat{f}^{\dagger}(x^{\prime},\omega^{\prime})] = \delta(x-x^{\prime})\delta(\omega - \omega^{\prime})$.  
The equal-time commutator $[\hat{\phi}(x),\hat{\rho}(y)] = \delta(x-y)$ follows from the above commutation relations and Green's identity\cite{supp}.\\\\
\textit{Complex Eigenvalues:} The surface impedance of a bulk superconductor depends on the complex conductivity, $\sigma(\omega) = \sigma_{1}(\omega) - i\sigma_{2}(\omega)$, as described by Mattis \& Bardeen\cite{MB}. Closed form expressions for $Z_{s}(\omega)$ are available for two thick-film limiting cases: the Extreme Anomalous limit for which the Cooper-pair coherence length $\zeta_{0}$ greatly exceeds the London penetration depth $\lambda_{L}$ \& the Dirty limit for which $\lambda_{L} >> \zeta_{0}$:
\begin{eqnarray}
    Z_{s}(\omega) \sim \begin{cases}
         \omega\left(\omega \frac{\sigma_{1}(\omega) - i\sigma_{2}(\omega)}{\sigma_{n}}\right)^{-1/3}, \;\;\;\; \text{Anomalous limit}  \\
         \omega \left(\omega \frac{\sigma_{1}(\omega) - i\sigma_{2}(\omega)}{\sigma_{n}}\right)^{-1/2}, \;\;\;\; \text{Dirty Limit}
    \end{cases}
\end{eqnarray}
Finding the complex fixed points of Eq. $10$, $\omega_{n} = \nu_{n} + i\kappa_{n}$ require evaluating the conductivity in the complex plane. We achieve this by an analytic continuation of the results of Mattis \& Bardeen, extending the conductivity to complex frequencies: $\sigma(\omega) \rightarrow \sigma(\omega,\kappa)$. Following the procedure by Mattis \& Bardeen, the conductivity is given by: $\sigma(\omega, \kappa) = \frac{I(\omega,\kappa)}{2\pi i \hbar(\omega + i\kappa)}$, where $I(\omega,\kappa)$ is the Mattis-Bardeen kernel in the spatial domain\cite{MB}. 
We proceed by evaluating this kernel for complex frequencies $\omega + i\kappa$. Within the Extreme Anomalous and Dirty limits, $I(\omega,\kappa)$ is determined by applying the residue theorem\cite{supp}  $I(\omega,\kappa) = \int d\epsilon d\epsilon^{\prime} L(\omega,\epsilon,\epsilon^{\prime})$, where the poles of $L(\omega,\epsilon,\epsilon^{\prime})$ satisfy:
\begin{eqnarray}
    \sqrt{\epsilon^{2} + \Delta^{2}} + \sqrt{\epsilon^{\prime 2} + \Delta^{2}} \pm \hbar(\omega + i\kappa) = 0
\end{eqnarray}
These poles depend on whether $\hbar\omega > 2\Delta$ or $\hbar\omega < 2\Delta$, leading to different behavior below and above the superconducting-gap. Since the estimated cut-off frequencies from Eq. $(1)$ are considerably greater than the superconducting-gaps of the commonly used metals in cQED, we proceed with the complex plane extension of the conductivity for the case of $\hbar\omega > 2\Delta$. This leads to the following expressions for the complex conductivity\cite{supp}:
\begin{eqnarray}
    \tilde{\sigma}_{1}(\nu,\kappa) &= \left(\frac{1}{2} + \frac{1}{\nu + i\kappa}\right)\left(E(e^{i\theta};k) + E(k)\right) \notag \\&\;\;\;\;\;\;\;\;\;\;\;\;\;\; - \frac{2}{\nu + i\kappa}\left(K(e^{i\theta}; k) + K(k)\right)
\end{eqnarray}
where $E(e^{i\theta},k)$ and $K(e^{i\theta},k)$ are the Incomplete Elliptic Integrals of the first and second kind, respectively. The phase $e^{i\theta} = \frac{\frac{\nu - i\kappa}{2} -1}{\frac{\nu + i\kappa}{2} -1}$ and argument $k=\frac{\frac{\nu + i\kappa}{2} - 1}{\frac{\nu + i\kappa}{2} + 1}$ , give $\theta = \frac{\pi}{2}$ for $\kappa =0$, thereby reducing the above formula to Mattis-Bardeen result. Similarly, $\sigma_{2}(\nu,\kappa)$ is given by: 
\begin{eqnarray}
    &&\frac{1}{2}\left(\frac{2}{\nu + i\kappa} + 1\right)\left\{E(k^{\prime}) - E\left(\frac{\sqrt{1-k_{1}^{2}k^{2}}}{k^{\prime}};k^{\prime}\right)\right\} \nonumber \\&&
     + \frac{1}{2}\left(\frac{2}{\nu + i\kappa} - 1\right)\left\{K(k^{\prime}) - K\left(\frac{\sqrt{1-k_{1}^{2}k^{2}}}{k^{\prime}};k^{\prime}\right)\right\}\nonumber
    \\
    &&+ \frac{i}{2}\left(\frac{2}{\nu + i\kappa} + 1\right)\left\{E(k) - E(e^{i\phi};k)\right\} \nonumber \\&& + \frac{2i}{\nu + i\kappa}\left\{K(k) - K(e^{i\phi};k)\right\}  
\end{eqnarray}
where the phase $e^{i\phi} = \frac{\frac{\nu + i\kappa}{2}-1}{\frac{\nu - i\kappa}{2}-1} $, argument $k_{1} = \frac{\frac{\nu + i\kappa}{2} + 1}{\frac{\nu - i\kappa}{2}-1}$, and the complimentary modulus $k^{\prime} = \sqrt{1-k^{2}}$. 
The expressions $(15,16)$ allow us to find $\{\nu_{n},\kappa_{n}\}$ by
fixed point iteration of Eq. (10), for both Aluminum (Extreme Anomalous Limit) and Niobium (Dirty Limit) resonators. 
\\\\
\textit{Light-Matter interaction:} 
For a non linear charge Qubit capacitively coupled to a resonator at position $x_{q}$, Eq. $(12)$ reveals that the coupling strength $g_{n}$ to the $n^{th}$ mode is given as:
\begin{eqnarray}
   g_{n} &&\propto \sqrt{\omega_{n}}\sqrt{\epsilon(\omega_{n})}\tilde{\Psi}_{n}(x_{q}) 
\end{eqnarray}
for $\omega_{n} \leq \frac{2\Delta}{\hbar}$; which gives way to an effective spectral density for frequencies beyond the superconducting-gap\cite{SD}:
\begin{eqnarray}
    J(\omega) &&\propto \omega^{2}\left(\frac{gR_{s}(\omega)}{\omega\ell_{m}}\right)Re\{\epsilon(\omega)\}Re\{G(x_{q},x_{q};\omega)\} \nonumber \\
&&\;\;\;\; + \omega^{2}|\epsilon(\omega)|^{2}Im\{G(x_{q},x_{q};\omega)\} 
\end{eqnarray}
where the proportionality is set by the dipole moment of the Qubit transition\cite{HakanA2}. 
\begin{figure}[h]
    \centering
    \includegraphics[width=\columnwidth]{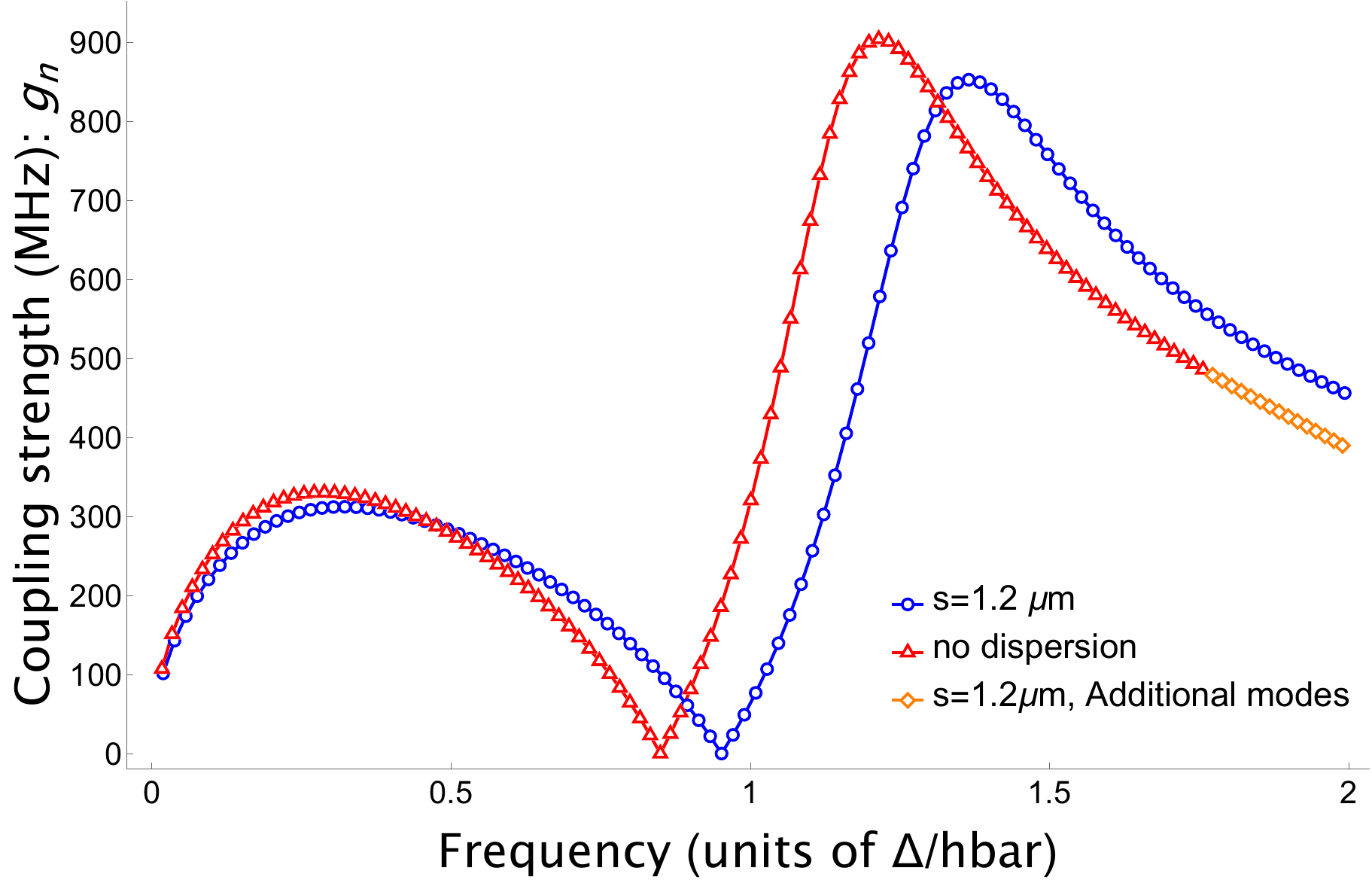}
    \caption{Coupling strengths $g_{n}$ vs Mode number $n$: For a Niobium resonator with center conductor width $= 1.2\mu m$, the modes are red-shifted in frequency due to dispersion. Close to the superconducting gap, additional modes (orange) are found due to large dispersive effects in Niobium.}
\end{figure}
We note that the total Lamb shift for a Two-Level System with frequency spacing $\Omega_{q}$ in a broadband environment is\cite{naturebroadbandLS, carmichael2013statistical}: $ \mathcal{P}\int_{0}^{\infty} d\omega \frac{J(\omega)}{\Omega_{q} - \omega} - \mathcal{P}\int_{0}^{\infty}\frac{J(\omega)}{\Omega_{q} + \omega} \equiv \mathcal{P}\int_{-\infty}^{\infty}\frac{J(\omega)}{\Omega_{q} -\omega}$, since $J(\omega) = -J(-\omega)$. We calculate the total Lamb shift $\Delta_{LS}$ using contour integration by closing in the upper half plane: 
\begin{eqnarray}
   \Delta_{LS} = &&(\pi i)\sum_{n}\frac{(\nu_{n} + i\kappa_{n})^{2}}{\Omega_{q} - \nu_{n} - i\kappa_{n}}\left(\frac{-\kappa_{n}}{\kappa_{n}^{2} + \nu_{n}^{2}}|\epsilon(\nu_{n} + i\kappa_{n})|^{2}\right. \nonumber \\
    && \left. + \frac{\nu_{n}}{\nu_{n}^{2} +\kappa_{n}^{2}} Re\{\epsilon(\nu_{n} + i\kappa_{n})\}Im\{\epsilon(\nu_{n} + i\kappa_{n})\} \right) \nonumber \\&& \times \Psi_{n}(x_{q})\Phi^{\star}_{n}(x_{q}) + \{\nu_{n} \rightarrow -\nu_{n}\}
\end{eqnarray}
\begin{figure}[h]
        \centering
        \includegraphics[width=\columnwidth]{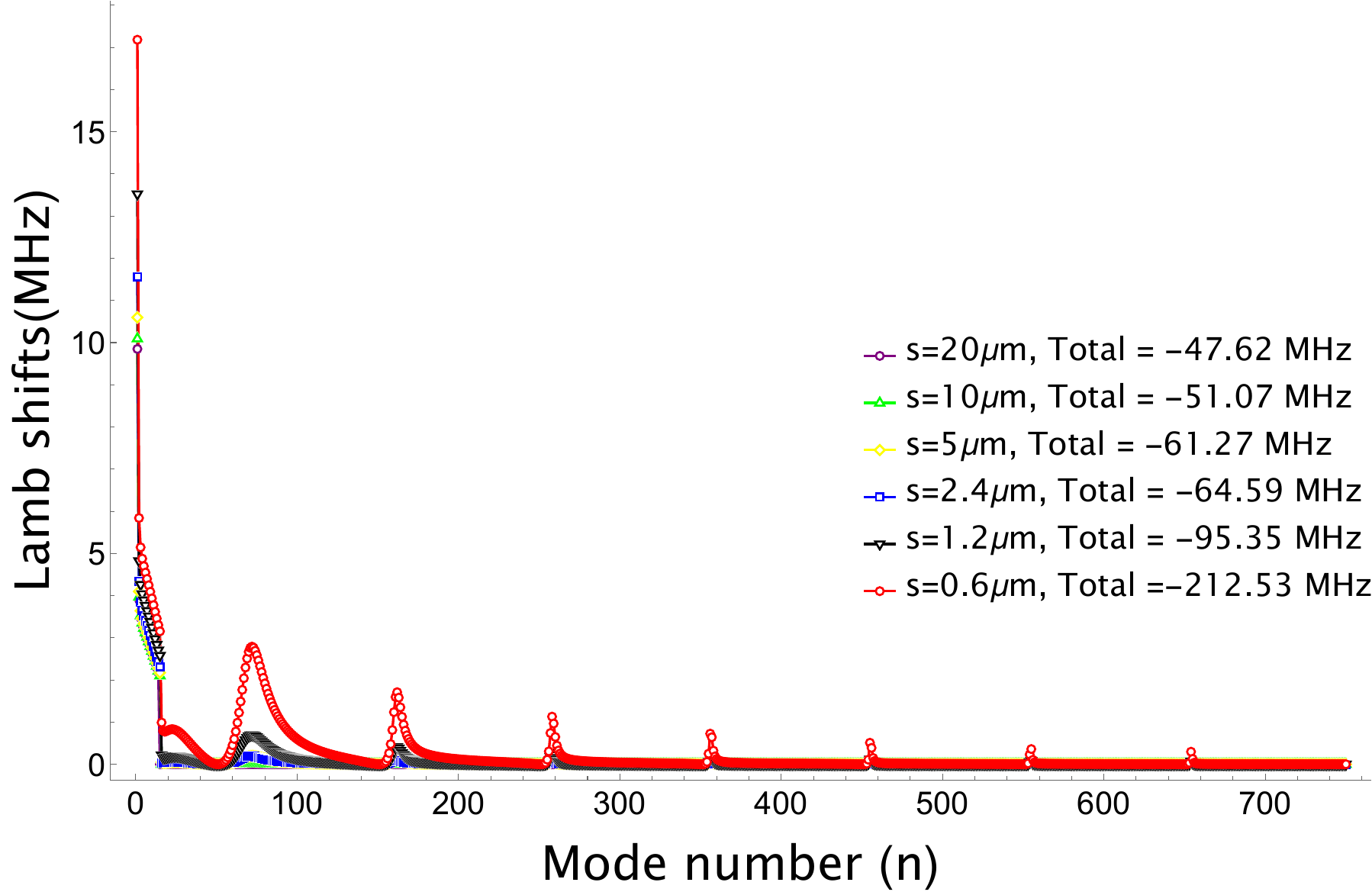}
        \caption{\label{fig3}Lamb shifts, obtained from the real part of the $n^{th}$ term in $\Delta_{LS}$, vs Mode number for varying center conductor width $s$ for a charge qubit coupled to a Aluminum Resonator.}
    \end{figure}
\begin{table*}[t!]
    \centering
\begin{tabular}{||c c c c c c||} 
 \hline
 $s(\mu m)$ & $f_{o}$(GHz) & $ g (m^{-1})\times 10^6$ & $\Delta_{dispersion}$(MHz) & $\Delta_{belowbandgap}$(MHz) & $\Delta_{no dispersion}$ (MHz)\\ [0.3ex] 
 \hline\hline
 0.6 &  6.93 & 3 &  -212.53 &  -76.26 & -272.612 \\ 
 \hline
 1.2 &  6.38 & 1.65  & -95.35 &  -62.29& -274.037 \\
 \hline
 2.4 & 6.14 & 0.9 & -64.59 &  -55.54& -274.97 \\
 \hline
 5 &  6.03 & 0.49 &  -61.27 &  -52.17 &-275.57\\
 \hline
 10 & 5.96 & 0.255 &  -51.07 &  -50.35 &-275.95 \\ 
 \hline 
 20 & 5.93 & 0.137 & -47.62 &  -47.41 &- 276.164\\ 
 \hline
\end{tabular}
    \caption{Parameters used for Fig. \ref{fig4}: $s$ is center conductor width, $f_{0}$ is the fundamental frequency of the resonator, $g$ is the geometric factor introduced in Eq. $(2)$, $\Delta_{dispersion}$ is the total Lamb shift accounting for the Surface Impedance. $\Delta_{belowbandgap}$ \& $\Delta_{nodispersion}$ are calculated from the expression for coupling strengths in \cite{HakanA2}. However, $\Delta_{belowbandgap}$ places a hard cut-off at the superconducting-gap. }
\end{table*}
Therefore, in Eq. $(19)$ the perturbative effect of the broadened modes has been mapped to a sum over the non-analytical points of the spectral density $J(\omega)$\cite{garraway, simplepole}.
We find that result $(19)$ provides both quantitatively different behavior depending on the width of the center conductor, Fig. \ref{fig3}.  For a single charge qubit coupled to a resonator with open boundary conditions, we find an order of magnitude drop in the Lamb-Shift contribution on increasing the center conductor width, suggesting that the cut-off strongly depends on the superconducting contribution. 
 \begin{figure}[h!]
        \centering
        \includegraphics[width=\columnwidth]{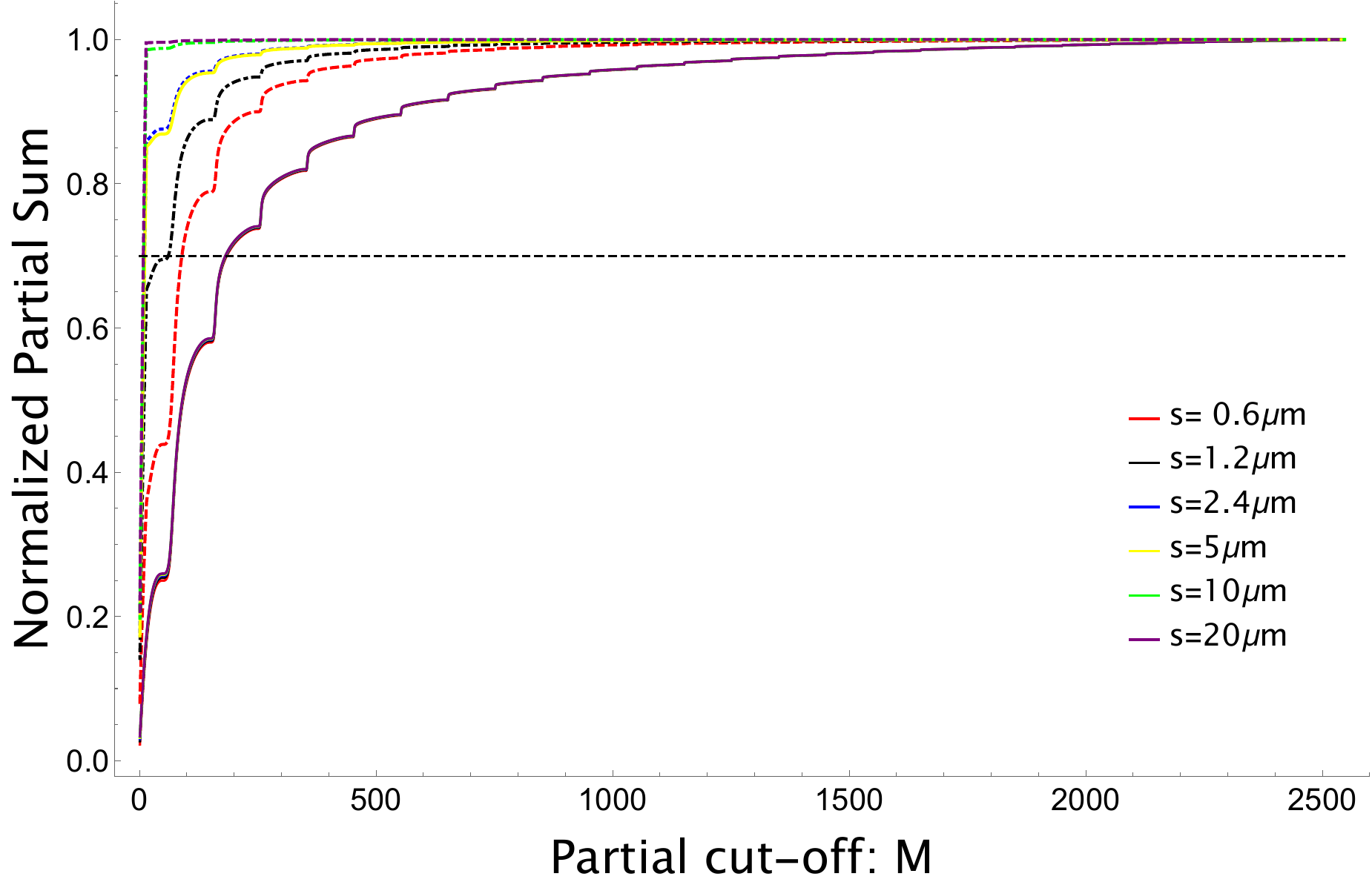}
        \caption{\label{fig4} Convergence of the normalized Lamb shift partial sums, $\frac{\sum_{n=1}^{M}\Delta_{LS}^{(n)}}{\sum_{n=1}^{2500}\Delta_{LS}^{(n)}}$ for varying widths $s$ as $M$ takes values between $1$ to $2500$. The solid curves, which are seen to be bunched together, are calculated using the expression for $g_{n}$ without any superconducting dispersion\cite{HakanA2}. The black dashed horizontal line denotes the $70 \%$ convergence.}
    \end{figure}\\\\
It is generally accepted in the Superconducting Circuit-QED community that modes beyond $2\Delta$ can be safely ignored in consideration of observables such as the Lamb shifts and inter-qubit coupling strengths. To investigate further, we set a high cut-off at mode number $n \sim 2500$ and consider the convergence of the Lamb shift partial series $\{\Delta_{LS}^{(n)}\}_{n=1}^{n=2500}$ in Fig. \ref{fig4}. The CC Hamiltonian\cite{HakanA2} provides universal convergence of the Lamb shift sums, independent of $s$. Due to the superconducting losses our results demonstrate a faster convergence rate which is inversely proportional to $s$. Finally, the contribution from beyond the superconducting gap modes is negligible only for the case of $s \geq 10 \mu m$. 
\\\\
\textit{Conclusions:} We presented an unified approach to model material-dependent effects by including the surface impedance of superconducting resonators in the canonical quantization procedure. Our work generalizes the cQED light matter interaction to all frequencies where the effects of dispersion cannot be ignored. To demonstrate the applicability of our method, we extended the Mattis-Bardeen complex conductivity to the complex plane and showed that due to superconducting loss the light \& matter degrees of freedom effectively decouple at high frequencies beyond the superconducting gap. We expect that our work holds relevance for improved modeling of multi-qubit superconducting systems, the development of which is currently impeded by losses originating from the superconducting material.   \\\\
\textit{Acknowledgments:} The authors thank David P. DiVincenzo and Athreya Shankar for their helpful comments and fruitful discussions. 
\bibliographystyle{apsrev4-2}
\bibliography{references}
\end{document}


\renewcommand{\theequation}{S\arabic{equation}}
\renewcommand{\thefigure}{S\arabic{figure}}
\renewcommand{\bibnumfmt}[1]{[S#1]}
\renewcommand{\citenumfont}[1]{S#1}
\title{\Large Supplementary Material: Light-Matter Interaction in dispersive Superconducting Circuit QED}
\author{Harsh Arora}
\affiliation{Center for Nano Science and Engineering, Indian Institute of Science, Bengaluru, India - 560012}
\author{Jay Deshmukh}
\affiliation{Department of Condensed Matter Physics, Tata Institute for Fundamenetal Research, Mumbai, India - 400005}
\author{Ansh Das}
\affiliation{Birla Institute of Technology and Science, Pilani, India - 333031}
\author{R. Vijay}
\affiliation{Department of Condensed Matter Physics, Tata Institute for Fundamenetal Research, Mumbai, India - 400005}
\author{Baladitya Suri}
\affiliation{Instrumentation and Applied Physics, Indian Institute of Science, Bengaluru, India - 560012}
\maketitle
\subsection*{\large Kramer Kronig Relations}
Time-causality implies that the  fourier transform $K(t) = \int_{-\infty}^{\infty}d\omega \left\{\tilde{\epsilon}(\omega)-1\right\}e^{i\omega t} $ satisfies $K(t) =0 $ for $t < 0$; implying that $\tilde{\epsilon}(\omega) - 1$ is analytic in the lower half plane. 
\begin{align*}
    \tilde{\epsilon}(\omega) - 1 = \frac{Im\{Z_{s}(\omega)\}}{\omega\ell_{m}} - i\frac{Re\{Z_{s}(\omega)\}}{\omega\ell_{m}}
\end{align*}
Then, on an infinitely large semi-circular contour closed in the lower half complex plane, we note that:
\begin{align}
    \oint d\omega
    \frac{\tilde{\epsilon}(\omega) - 1}{\omega - \omega^{\prime} + i\epsilon} = -2\pi i\left(\tilde{\epsilon}(\omega^{\prime} - i \epsilon)-1\right)
\end{align}
where the integral is taken in the clock wise direction. Separating the above integral into contributions along the real axis and the semi-circular arc of infinite radius, we get: 
\begin{eqnarray}
    &&\oint d\omega \frac{\tilde{\epsilon}(\omega) - 1}{\omega - \omega^{\prime}} \nonumber \\ &=& \int_{I} d\omega \frac{\tilde{\epsilon}(\omega) - 1}{\omega - \omega^{\prime}} + \lim_{R\rightarrow \infty}R\int_{0}^{-\pi}  id\theta e^{i\theta}\frac{\tilde{\epsilon}(R e^{i\theta})-1}{Re^{i\theta} - \omega^{\prime}} \nonumber\\
    &=& \int_{I} d\omega \frac{\tilde{\epsilon}(\omega) - 1}{\omega - \omega^{\prime}} + i\int_{0}^{-\pi} d\theta \left(\tilde{\epsilon}(R e^{i\theta}) - 1\right) \nonumber \\
    &=& \int_{I}d\omega \frac{\tilde{\epsilon}(\omega) - 1}{\omega - \omega^{\prime}} \nonumber
\end{eqnarray}
where the contribution from the semi-circular arc goes to zero in the limit of $R \rightarrow \infty$, since: $|Z_{s}(\omega)| \sim \omega^{2/3}$ as $\omega \rightarrow \infty$. The contribution from the deformed real axis is:
\begin{align*}
    &\mathcal{P}\int_{-\infty}^{\infty} d\omega \frac{\tilde{\epsilon}(\omega)-1}{\omega - \omega^{\prime}} + \lim_{\rho \rightarrow 0^{+}}\int_{\pi}^{0}d\theta i\rho e^{i\theta}\frac{\tilde{\epsilon}(\omega^{\prime} + \rho e^{i\theta})-1}{\rho e^{i\theta}} \\
    &= \mathcal{P}\int_{-\infty}^{\infty} d\omega \frac{\tilde{\epsilon}(\omega)-1}{\omega - \omega^{\prime}} - i\pi \left(\tilde{\epsilon}(\omega^{\prime}) - 1\right)
\end{align*}
where the integration contour is slightly bumped up at $\omega^{\prime}$. Finally, we separate the above contribution into real and imaginary parts and equate the result to $(1)$, and take the limit $\epsilon \rightarrow 0^{+}$:
\begin{eqnarray}
    \mathcal{P}\int_{-\infty}^{\infty}\frac{Im\{Z_{s}(\omega)\}}{\omega(\omega - \omega^{\prime})} - i\mathcal{P}\int_{-\infty}^{\infty}\frac{Re\{Z_{s}(\omega)\}}{\omega(\omega - \omega^{\prime})} \nonumber \\ + i\pi \frac{Im\{Z_{s}(\omega)\}}{\omega} + \pi\frac{Re\{Z_{s}(\omega)\}}{\omega} = 0 \notag 
\end{eqnarray}
which gives the Kramer-Kronig relations:
\begin{eqnarray}
    \mathcal{P}\int_{-\infty}^{\infty}d\omega \frac{Re\{Z_{s}(\omega)\}}{\omega(\omega - \omega^{\prime})} &=& \pi \frac{Im\{Z_{s}(\omega^{\prime})\}}{\omega^{\prime}} \\
    \mathcal{P}\int_{-\infty}^{\infty}d\omega \frac{Im\{Z_{s}(\omega)\}}{\omega(\omega - \omega^{\prime})} &=& -\pi \frac{Re\{Z_{s}(\omega^{\prime})\}}{\omega^{\prime}}
\end{eqnarray}
We can further simplify the above relations, to give:
\begin{eqnarray}
\mathcal{P}\int_{0}^{\infty}d\omega \frac{Re\{Z_{s}(\omega)\}}{\omega^{2} - \omega^{\prime 2}} = \frac{\pi}{2}\frac{Im\{Z_{s}(\omega^{\prime})\}}{\omega^{\prime}}
\end{eqnarray}
    The same relations hold for the dirty limit, relevant for the case of Niobium as well since $|Z_{dirty}(\omega)| \sim \omega^{1/2}$ in the limit of $\omega \rightarrow \infty$. 
\onecolumngrid
\subsection*{\large Generalization of the Current Conserving Lagrangian to include $Z_{s}(\omega)$} 
We begin with a discrete version of the total Lagrangian density and derive the equations central to the results: namely, wave-equation $(1)$, the Hamiltonian $(5)$ and equations $(7)$ \& $(8)$. The analysis presented here is an extension of work presented previously by different authors\cite{HakanA2}. By taking the continuum limit of the discrete Lagrangian we arrive at the Lagrangian density provided in the main text.  
\begin{eqnarray}
    L &=& \frac{c\Delta x}{2}\sum_{n}\dot{\phi}_{n}^{2} - \frac{1}{2\ell_{m}\Delta x}\sum_{n}\left(\phi_{n} - \phi_{n+1}\right)^{2} + \sum_{i}\frac{C_{J_{i}}}{2}\dot{\Phi}_{J_{i}}^{2} + E_{J_{i}}\cos\left(\frac{2\pi}{\phi_{0}}\Phi_{J_{i}}\right) 
     + \sum_{i}\frac{C_{g_{i}}}{2}\left(\dot{\phi_{i}} - \dot{\Phi}_{J_{i}}(t)\right)^{2} \notag \\ && + \sum_{n}\Delta x\int_{0}^{\infty}d\omega \left(\frac{\partial \psi_{\omega}(x_{n})}{\partial t}\right)^{2} - \omega^{2}\psi^{2}_{\omega}(x_{n})  -  \sqrt{\frac{2}{\pi}}\sum_{n}\Delta x\int_{0}^{\infty} d\omega \sqrt{\frac{gRe\{Z_{s}(\omega)\}c(x_{n})}{\ell_{m}}}\psi_{\omega}(x_{n})\dot{\phi}_{n}
\end{eqnarray}
where the sum over all the transmons and the nodes $n$ is included. The current conservation conditions: 1. discontinuity of the flux derivative at the transmon location \& 2. continuity of the flux $\phi(x,t)$ arise due to the modification of the capacitance per unit length of the transmission line from $c \rightarrow c(x_{n}) = c + C_{s,j}\delta_{n,j}$. This modification only arises if we retain the di-magnetic term. Note that the interaction between the reservoir and transmission line contains the modified capacitance per unit length from the beginning. The conjugate momenta are given by: 
\begin{eqnarray}
    \rho_{n}  &=& \frac{\partial L}{\partial \dot{\phi}_{n}} = (c\Delta x + \sum_{j}\delta_{n,j}C_{g_{j}})\dot{\phi}_{n} - \sum_{j}\delta_{n,j}C_{g_{j}}\dot{\Phi}_{J_{j}}  - \sqrt{\frac{2}{\pi}}\Delta x\int_{0}^{\infty}d\omega \sqrt{\frac{gRe\{Z_{s}(\omega)\}c(x_{n})}{\ell_{m}}}\psi_{\omega}(x_{n}) \\
    Q_{J_{j}} &=& \frac{\partial L}{\partial \dot{\phi}_{J_{j}}} = (C_{g_{j}} + C_{J_{j}})\dot{\Phi}_{J_{j}} - C_{g_{j}}\dot{\phi}_{j} \\
    \rho_{\omega,n} &=& \frac{\partial L}{\partial \dot{\psi}_{\omega}(x_{n})} = \Delta x\dot{\psi}_{\omega}(x_{n})
\end{eqnarray}
To find the Hamiltonian we need to invert the above equations to find the generalized velocities in terms of conjugate momenta: 
\begin{eqnarray}
    \dot{\phi}_{n} &=& \frac{\rho_{n}}{c\Delta x + \sum_{j}\delta_{n,j}C_{s,j}} + \Delta x\sqrt{\frac{2}{\pi}}\frac{\int_{0}^{\infty} d\omega \sqrt{\frac{gRe\{Z_{s}(\omega)\}c(x_{n})}{\ell_{m}}}\psi_{\omega}(x_{n}) }{c\Delta x + \sum_{j}\delta_{n,j}C_{s,j}} + \frac{\sum_{j}\delta_{n,j}\gamma_{j}Q_{J_{j}}}{c\Delta x + \sum_{j}\delta_{n,j}C_{s,j}} \\
    \dot{\Phi}_{J_{j}} &=& \frac{Q_{J_{j}}}{C_{g_{j}} + C_{J_{j}}} + \gamma_{j}\frac{\rho_{j}}{c\Delta x + C_{s,j}} + \gamma_{j}\Delta x\sqrt{\frac{2}{\pi}}\frac{\int_{0}^{\infty} d\omega \sqrt{\frac{gRe\{Z_{s}(\omega)\}c(x_{n})}{\ell_{m}}}\psi_{\omega}(x_{j}) }{c\Delta x + C_{s,j}}  + \gamma_{j}^{2}\frac{Q_{J_{j}}}{c\Delta x + C_{s,j}}
\end{eqnarray}
where similar to Ref\cite{HakanA2} we have defined the dimensionless constant $\gamma_{j} = \frac{C_{g,j}}{C_{g,j} + C_{J,j}}$ and the series capacitance $C_{s,j} = \frac{C_{g_{j}}C_{J_{j}}}{C_{g_{j}} + C_{J_{j}}}$. \\\\
The Hamiltonian is given by the Legendre transform: 
\begin{align}
    H = \sum_{n}\rho_{n}\dot{\phi}_{n} + \sum_{j}Q_{J_{j}}\dot{\Phi}_{J_{j}} + \sum_{n}\rho_{\omega,n}\dot{\psi}_{\omega}(x_{n}) - L 
\end{align}
The inverted equations for the generalized velocities (voltages) contain additional contributions from the reservoir modes. This implies that the Hamiltonian also will contain additional terms which will correspond to couplings to the reservoir modes, as expected. However we can greatly simplify the analysis by noting that the terms which do not depend on the reservoir modes essentially transform and appear in the Hamiltonian in exactly the same manner as when the reservoir modes were completely absent. Therefore, in the following equations we only present the additional terms that arise due to the coupling to the reservoir. We consider each of the terms in equation $(11)$ and the terms in $-L$:    
\begin{eqnarray}
    \rho_{n}\dot{\phi}_{n} &=&   \rho_{n}\Delta x\sqrt{\frac{2}{\pi}}\frac{\int_{0}^{\infty} d\omega \sqrt{\frac{gRe\{Z_{s}(\omega)\}c(x_{n})}{\ell_{m}}}\psi_{\omega}(x_{n}) }{c\Delta x + \sum_{j}\delta_{n,j}C_{s,j}} \\
    Q_{J_{j}}\dot{\phi}_{J_{j}} &=& Q_{J_{j}}\gamma_{j}\Delta x\sqrt{\frac{2}{\pi}}\frac{\int_{0}^{\infty} d\omega \sqrt{\frac{gRe\{Z_{s}(\omega)\}c(x_{n})}{\ell_{m}}}\psi_{\omega}(x_{j}) }{c\Delta x + C_{s,j}}
\end{eqnarray}
The final term $\rho_{\omega,n}\dot{\psi}_{\omega}(x_{n})$ is straightforward: $\dot{\psi}_{\omega}(x_{n}) = \frac{\rho_{\omega,n}}{\Delta x}$. Turning to the terms from $-L$, we first consider the Kinetic Energy terms corresponding to $\dot{\phi}_{n}^{2}$, $\dot{\Phi}_{J}^{2}$ and the coupling $\dot{\phi}_{n}\dot{\Phi}_{J}$: 
\begin{eqnarray}
    -\frac{1}{2}(c\Delta x + \sum_{j}\delta_{n,j}C_{g_{j}})\dot{\phi}_{n}^{2} &=& \left( -\frac{\rho_{n}}{(c\Delta x + \sum_{j}\delta_{n,j}C_{s,j})}  -\frac{\sum_{j}\delta_{n,j}\gamma_{j}Q_{J_{j}}}{c\Delta x + \sum_{j}\delta_{n,j}C_{s,j}} - \frac{1}{2}(\Delta x)\sqrt{\frac{2}{\pi}}\frac{\int_{0}^{\infty} d\omega \sqrt{\frac{gRe\{Z_{s}(\omega)\}c(x_{n})}{\ell_{m}}}\psi_{\omega}(x_{n})}{c\Delta x + \sum_{j}\delta_{n,j}C_{s,j}}\right) \nonumber \\ && \;\;\;\;\;\; \times \Delta x\sqrt{\frac{2}{\pi}}\frac{\int_{0}^{\infty} d\omega \sqrt{\frac{gRe\{Z_{s}(\omega)\}c(x_{n})}{\ell_{m}}}\psi_{\omega}(x_{n}) }{(c\Delta x + \sum_{j}\delta_{n,j}C_{s,j})}(c\Delta x + \sum_{j}\delta_{n,j}C_{g_{j}})
\end{eqnarray}
\begin{eqnarray}
    -\frac{C_{J_{j}} + C_{g_{j}}}{2}\dot{\Phi}_{J_{j}}^{2} &=& -\frac{C_{g_{j}} + C_{J_{j}}}{2}\left((\Delta x)\gamma_{j}^{2}\sqrt{\frac{2}{\pi}}\frac{\int_{0}^{\infty} d\omega \sqrt{\frac{gRe\{Z_{s}(\omega)\}c(x_{n})}{\ell_{m}}}\psi_{\omega}(x_{j})}{c\Delta x + C_{s,j}} + 2\frac{Q_{J_{j}}}{C_{g_{j}} + C_{J_{j}}}\gamma_{j}  + 2\frac{\rho_{j}}{c\Delta x + C_{s_{j}}}\gamma_{j}^{2}\right. \notag \\
    &&\;\;\;\;\;\;\;\;\; \left. +2\gamma_{j}^{3}\frac{Q_{J_{j}}}{c\Delta x + C_{s,j}}\right)\times \Delta x\sqrt{\frac{2}{\pi}}\frac{\int_{0}^{\infty} d\omega \sqrt{\frac{gRe\{Z_{s}(\omega)\}c(x_{n})}{\ell_{m}}}\psi_{\omega}(x_{j}) }{c\Delta x + C_{s,j}} 
\end{eqnarray}
\begin{eqnarray}
    C_{g_{j}}\dot{\Phi}_{J_{j}}\dot{\phi}_{j} &=& C_{g_{j}}\left(2\frac{\rho_{j}}{c\Delta x + C_{s,j}}\gamma_{j}  + \frac{Q_{J_{j}}}{C_{g_{j}} + C_{J_{j}}}   +  (\Delta x)\gamma_{j}\sqrt{\frac{2}{\pi}}\frac{\int_{0}^{\infty} d\omega \sqrt{\frac{gRe\{Z_{s}(\omega)\}c(x_{n})}{\ell_{m}}}\psi_{\omega}(x_{j})}{c\Delta x + C_{s,j}} \right) \nonumber \\
    &&\left. \;\;\;\;\;\;\;\; + \gamma_{j}\frac{\gamma_{j}Q_{J_{j}}}{c\Delta x + C_{s,j}} +\gamma_{j}^{2}\frac{Q_{J_{j}}}{c\Delta x + C_{s,j}}\right) \Delta x\sqrt{\frac{2}{\pi}}\frac{\int_{0}^{\infty} d\omega \sqrt{\frac{gRe\{Z_{s}(\omega)\}c(x_{n})}{\ell_{m}}}\psi_{\omega}(x_{j}) }{c\Delta x + C_{s,j}}
\end{eqnarray}
The contribution from the last term in Eq. $(S5)$ gives:
\begin{eqnarray}
    &&\sqrt{\frac{2}{\pi}}\Delta x\int_{0}^{\infty} d\omega \sqrt{\frac{gRe\{Z_{s}(\omega)\}c(x_{n})}{\ell_{m}}}\psi_{\omega}(x_{n})\dot{\phi}_{n} \nonumber \\  && = \left(\frac{\rho_{n}}{c\Delta x + \sum_{j}\delta_{n,j}C_{s,j}}  + \frac{\sqrt{\frac{2}{\pi}}\sum_{n}\Delta x\int_{0}^{\infty} d\omega \sqrt{\frac{gRe\{Z_{s}(\omega)\}c(x_{n})}{\ell_{m}}}\psi_{\omega}(x_{n})}{c\Delta x + \sum_{j}\delta_{n,j}C_{s,j}} -\frac{\sum_{j}\delta_{n,j}\gamma_{j}Q_{J_{j}}}{c\Delta x + \sum_{j}\delta_{n,j}C_{s,j}}\right)\nonumber \\ &&\;\;\;\;\;\;\;\;\;\;\; \times \sqrt{\frac{2}{\pi}}\Delta x\int_{0}^{\infty} d\omega \sqrt{\frac{gRe\{Z_{s}(\omega)\}c(x_{n})}{\ell_{m}}}\psi_{\omega}(x_{n})
\end{eqnarray}
The potential energy terms do not undergo any change. From the above equations we note that the charge on the transmon, $Q_{J}$ and the charge density on the resonator $\rho_{n}$ is coupled to the reservoir mode $\psi_{\omega}$. We sum equations $(12)-(17)$ and find that the coupling between the reservoir and resonator charge is given as: 
\begin{eqnarray}
    \frac{\rho_{n}}{c\Delta x + \sum_{n,j}C_{s,j}}\sqrt{\frac{2}{\pi}}\Delta x\int_{0}^{\infty} d\omega \sqrt{\frac{gR_{s}(\omega)c(x_{n})}{\ell_{m}}} \psi_{\omega}(x_{n})
\end{eqnarray}
Similarly, we find that the coupling between the transmon charge $Q_{J_{j}}$ and reservoir modes is zero:
\begin{eqnarray}
    &&\frac{\gamma_{j}(C_{g,j} - C_{s,j}) + \gamma_{j}^{3}(C_{g,j} + C_{J,j}) -2\gamma_{j}^{2}C_{g,j}}{(c\Delta x + C_{s,j})^{2}} \nonumber \\ &=& \frac{(\gamma_{j}-1)^{2}\gamma_{j}C_{g,j} - \gamma_{j}C_{s,j} + \gamma_{j}^{3}C_{J,j}}{(c\Delta x + C_{s,j})^{2}}\nonumber  \\
    &=& \frac{1}{(c\Delta x + C_{s,j})^{2}}\left(\frac{C_{J_{j}}^{2}}{(C_{g_{j}} + C_{J_{j}})^{2}}\frac{C_{g_{j}}^{2}}{(C_{g_{j}} + C_{J_{j}})} - \gamma_{j}^{2}C_{J_{j}} + \gamma_{j}^{3}C_{J,j}\right) \nonumber\\ 
    &=&\frac{1}{(c\Delta x + C_{s,j})^{2}}\left(\gamma_{j}^{2}\frac{C_{J_{j}}^{2}}{C_{g_{j}} + C_{J_{j}}} - \gamma_{j}^{2}C_{J_{j}} + \gamma_{j}^{3}C_{J,j}\right) \nonumber \\
    &=& 0
\end{eqnarray}
Finally, we have the coefficient for $\psi_{\omega}^{2}$ which corresponds to the reservoir potential energy:
\begin{eqnarray}
    &&\frac{1}{(c\Delta x + \sum_{n,j}\delta_{n,j}C_{s,j})^{2}}\left(-\frac{c\Delta x + \sum_{n,j}\delta_{n,j}C_{g_{j}}}{2} - \frac{1}{2}\frac{C_{g_{j}}^{2}}{C_{g_{j}} + C_{J_{j}}} + \frac{C_{g_{j}}^{2}}{C_{g_{j}} + C_{J_{j}}} + (c\Delta x + \sum_{j}\delta_{n,j}C_{s,j})\right) \nonumber \\ &=& \frac{1}{2}\frac{1}{c\Delta x + \sum_{n}\delta_{n,j}C_{s,j}}
\end{eqnarray}
Results $(S18)-(S20)$ show that including the surface impedance $Z_{s}(\omega)$ to capture dispersion leads to a charge density-reservoir interaction, while leaving the qubit-resonator interaction proportional to the product of the charges, as is expected from the current conserving Hamiltonian. Therefore, the generalization of the current conserving Hamiltonian is complete. To proceed to the continuous limit we define the charge density: $\rho(x) = \lim_{\Delta x \rightarrow 0}\frac{\rho(x_{n})}{\Delta x}$ and dirac delta $\delta(x-x_{j}) = \lim_{\Delta x \rightarrow 0}\frac{\delta_{n,j}}{\Delta x}$, giving us $H = H_{TL,R} + H_{Transmon} + H_{int}$: 
\begin{eqnarray}
 H_{TL,R} &&=  \frac{1}{2}\int_{0}^{L}dx\frac{1}{c(x)}\left(\rho(x) + \sqrt{\frac{2}{\pi}}\int_{0}^{\infty}d\omega \sqrt{\frac{gRe\{Z_{s}(\omega)\}c(x)}{\ell_{m}}}\psi_{\omega}(x)\right)^{2} + \frac{1}{2\ell_{m}}\int_{0}^{L}dx \left(\frac{\partial \phi}{\partial x}\right)^{2}  \nonumber \\ &&  + \frac{1}{2}\int_{0}^{\infty} d\omega \int_{0}^{L}dx \rho_{\omega}^{2}(x) + \omega^{2}\psi_{\omega}^{2}(x) \\
 H_{Transmon} &&= \sum_{j}\frac{1}{2C_{J_{j}}}Q_{J_{j}}^{2} - E_{J_{j}}\cos\left(2\pi\frac{\Phi_{J_{j}}}{\phi_{o}}\right) \\
    H_{int} &&= \sum_{j}\gamma_{j}Q_{J_{j}}\int_{0}^{L} dx \frac{\rho(x,t)\delta(x-x_{j})}{c(x)}
\end{eqnarray}
where the Hamiltonian $H_{TL,R}$ contains the Resonator-Reservoir interaction. The above decomposition of $H$ makes it clear that we must proceed with the normal modes of the $H_{TL,R}$, which are solutions of the equations of motion obtained either by the Hamilton equations or Euler-Lagrange equations by only considering the Hamiltonian $H_{TL,R}$.     
\subsection*{\large Fano Diagonalization: Dressed modes of the Resonator obtained from $\mathcal{H}_{TL,R}$}
The equations of motion are given by: 
\begin{eqnarray}
    \frac{\partial \phi}{\partial t} &&= \frac{1}{c(x)}\left(\rho(x,t) + \sqrt{\frac{2}{\pi}}\int_{0}^{\infty}d\omega \sqrt{\frac{gRe\{Z_{s}(\omega)\}c(x)}{\ell_{m}}}\psi_{\omega}(x)\right) \\
    \frac{\partial \rho_{\omega}}{\partial t} &&= - \omega^{2}\psi_{\omega}(x,t) - \frac{1}{c(x)}\left(\rho(x,t) + \sqrt{\frac{2}{\pi}}\int_{0}^{\infty}d\omega^{\prime} \sqrt{\frac{gRe\{Z_{s}(\omega^{\prime})\}c(x)}{\ell_{m}}}\psi_{\omega^{\prime}}(x)\right)\sqrt{\frac{gRe\{Z_{s}(\omega)\}c(x)}{\ell_{m}}} \\
    \frac{\partial \rho}{\partial t} &&= \frac{1}{\ell_{m}}\frac{\partial^{2} \phi}{\partial x^{2}} \\
    \frac{\partial \psi_{\omega}}{\partial t} &&= \rho_{\omega}(x,t) \\
\end{eqnarray}
Eliminating the charge density $\rho(x,t)$ and reservoir momenta $\rho_{\omega}(x,t)$ gives us the following equations (equivalently these can be obtained directly from Lagrangian $(S5)$):
\begin{eqnarray}
    \ell_{m}c(x)\frac{\partial^{2}\phi}{\partial t^{2}} - \frac{\partial^{2}\phi}{\partial x^{2}} - \sqrt{\frac{2}{\pi}}\int_{0}^{\infty}d\omega \sqrt{g\ell_{m}Re\{Z_{s}(\omega)\}c(x)}\dot{\psi}_{\omega}(x) &&= 0 \\
    \frac{\partial^{2}\psi_{\omega}(x)}{\partial t^{2}} + \omega^{2}\psi_{\omega}(x) + \sqrt{\frac{2}{\pi}}\sqrt{\frac{gRe\{Z_{s}(\omega)\}c(x)}{\ell_{m}}}\frac{\partial \phi}{\partial t} &&= 0
\end{eqnarray}
We express the above equations in the Fourier domain via the following relations:
\begin{align*}
    \psi_{\omega}(x,t) &= \int_{-\infty}^{\infty} d\omega^{\prime} \tilde{\psi}_{\omega}(x,\omega^{\prime})e^{i\omega^{\prime}t} = \int_{0}^{\infty} d\omega^{\prime} \tilde{\psi}_{\omega}(x,\omega^{\prime})e^{i\omega^{\prime}t} + \int_{0}^{\infty}d\omega^{\prime} \tilde{\psi}_{\omega}^{\star}(x,\omega^{\prime})e^{-i\omega^{\prime}t} \\
    \phi(x,t) &= \int_{-\infty}^{\infty} d\omega^{\prime} \tilde{\phi}(x,\omega^{\prime})e^{i\omega^{\prime} t} = \int_{0}^{\infty}d\omega^{\prime} \tilde{\phi}(x,\omega^{\prime})e^{i\omega^{\prime} t} + \int_{0}^{\infty}d\omega^{\prime}\tilde{\phi}^{\star}(x,\omega^{\prime})e^{-i\omega^{\prime} t} 
\end{align*}
where the following symmetry of the solutions has been utilised: $\tilde{\phi}(x,-\omega) = \tilde{\phi}^{\star}(x,\omega)$, similarly for the $\tilde{\psi}_{\omega}$. Then, the equation of motion for the bath field $(S30)$, expressed in Fourier domain leads to:
\begin{align}
    -\omega^{\prime 2}\psi_{\omega}(x,\omega^{\prime}) + \omega^{2}\psi_{\omega}(x,\omega^{\prime}) + \sqrt{\frac{2}{\pi}}\sqrt{\frac{gRe\{Z_{s}(\omega)\}c(x)}{\ell_{m}}}(i\omega^{\prime})\tilde{\phi}(x,\omega^{\prime}) = 0
\end{align}
The solution of the above equation can be written as a sum of a homogeneous solution and a particular solution, giving: 
\begin{align}
    \psi_{\omega}(x,\omega^{\prime}) = -\sqrt{\frac{2}{\pi}}\sqrt{\frac{gRe\{Z_{s}(\omega)\}c(x)}{\ell_{m}}} \frac{(i\omega^{\prime})\tilde{\phi}(x,\omega^{\prime})}{\omega^{2} - \omega^{\prime 2}} + \delta(\omega - \omega^{\prime})\mathbf{X}_{\omega}(x) + \delta(\omega + \omega^{\prime})\mathbf{X}^{\star}_{\omega}(x)
\end{align}
where the solutions $\mathbf{X}_{\omega}(x)$ and it's complex conjugate satisfy the equation of motion $(S31)$ with $\phi = 0$. This amounts to $\mathbf{X}_{\omega}$ being associated with the positive frequency component of $\psi_{\omega}(x,t)$, since the homogeneous solution in time-domain is given by: 
\begin{align}
    \psi_{\omega}(x,t) = \mathbf{X}_{\omega}(x)e^{i\omega t} + \mathbf{X}^{\star}_{\omega}(x)e^{-i\omega t}
\end{align}

Similarly, the equation of motion for the node flux expressed in frequency domain is:
\begin{align*}
    -\omega^{\prime 2}\ell_{m}c(x)\tilde{\phi}(x,\omega^{\prime}) - \frac{\partial^{2}\tilde{\phi}(x,\omega^{\prime})}{\partial x^{2}} - \sqrt{\frac{2}{\pi}}\int d\omega\sqrt{g\ell_{m}Re\{Z_{s}(\omega)\}c(x)}(i\omega^{\prime})\tilde{\psi}_{\omega}(x,\omega^{\prime}) = 0
\end{align*}
Plugging in the solution $(S32)$ for $\psi_{\omega}(x,\omega^{\prime})$, we get a sum of two contributions, one from the particular solution and the other from the homogeneous solution. The particular solution gives: 
\begin{align*}
&\frac{2}{\pi}\int d\omega \sqrt{\ell_{m}Re\{Z_{s}(\omega)\}c(x)}(i\omega^{\prime})\frac{(i\omega^{\prime})\sqrt{\frac{Re\{Z_{s}(\omega)\}c(x)}{\ell_{m}}}\tilde{\phi}(x,\omega^{\prime})}{\omega^{2} - \omega^{\prime 2}} \\ 
     &= -\frac{2}{\pi}\omega^{\prime 2}c(x)\tilde{\phi}(x,\omega^{\prime})\int_{0}^{\infty} d\omega \frac{Re\{Z_{s}(\omega)\}}{\omega^{2} - \omega^{\prime 2}} \\
     &=  -\frac{2}{\pi}\omega^{\prime 2}c(x)\tilde{\phi}(x,\omega^{\prime})\left(\mathcal{P}\int_{0}^{\infty} d\omega \frac{Re\{Z_{s}(\omega)\}}{\omega^{2} - \omega^{\prime 2}} + \lim_{\rho \rightarrow 0^{+}}\int_{\pi}^{0} d\theta (i\rho e^{i\theta})\frac{Re\{Z_{s}(\omega^{\prime} + \rho e^{i\theta})\}}{(\rho e^{i\theta})(2\omega^{\prime} + \rho e^{i\theta})} \right) \\
     &=  -\frac{2}{\pi}\omega^{\prime 2}c(x)\tilde{\phi}(x,\omega^{\prime})g\left(\frac{\pi}{2}\frac{Im\{Z_{s}(\omega^{\prime})\}}{\omega^{\prime}} - i\frac{\pi}{2}\frac{Re\{Z_{s}(\omega^{\prime})\}}{\omega^{\prime}}\right) 
\end{align*}
where, as done previously, we deform the integration contour at $\omega^{\prime}$ so that the singularity at $\omega^{\prime}$ is included and employed the Kramer-Kronig relations $(S4)$. The homogeneous part gives us:
\begin{align*}
    &-\sqrt{\frac{2}{\pi}}\int_{0}^{\infty} d\omega \sqrt{g\ell_{m}Re\{Z_{s}(\omega)\}c(x)}(i\omega^{\prime})\left(\delta(\omega - \omega^{\prime})\mathbf{X}_{\omega^{\prime}}(x) + \delta(\omega + \omega^{\prime})\mathbf{X}_{\omega^{\prime}}^{\star}\right)
\end{align*}
It is important here to note that $\omega$, which denotes the energy of a bath mode, takes strictly positive values. However, $\omega^{\prime}$ denotes the Fourier frequency of the functions $\phi(x,t)$ and $\psi_{\omega}(x,t)$, therefore takes values from $-\infty$ to $\infty$. Hence, both parts of the homogeneous solutions contribute, with $\mathbf{X}_{\omega^{\prime}}$ contributing for $\omega^{\prime} > 0$ and $\mathbf{X}^{\star}_{\omega^{\prime}}$ contributing for $\omega^{\prime} <0$. \\\\
Therefore, the equation of motion for $\tilde{\phi}(x,\omega^{\prime})$ satisfies:
\begin{eqnarray}
    \omega^{\prime 2}\ell_{m}c(x)\epsilon(\omega)\tilde{\phi}(x,\omega^{\prime}) + \frac{\partial^{2}\tilde{\phi}(x,\omega^{\prime})}{\partial x^{2}} = -i\sqrt{g\ell_{m}R_{s}(\omega)c(x)}\omega^{\prime}\mathbf{X}_{\omega^{\prime}}(x)
\end{eqnarray}
which is the inhomogeneous wave equation for the frequency interval where $Re\{Z_{s}(\omega^{\prime})\} \neq 0$, which corresponds to $\omega^{\prime} > \frac{2\Delta}{\hbar}$. The complex conjugate of the above equation is satisfied by $\tilde{\phi}^{\star}(x,\omega^{\prime})$, which is equivalent to solving the equation of motion for $\tilde{\phi}(x,-\omega^{\prime})$.  The frequency-dependent refractive index is defined as: $\epsilon(\omega) =  1 + \frac{g}{i\omega \ell_{m}}Z_{s}(\omega)$. Since we assume zero temperature, the classical limit is obtained by setting $\mathbf{X}_{\omega}(x) =0$ which leads to the non-hermitian wave equation $(2)$ in the main text. For frequencies $\hbar\omega^{\prime} < 2\Delta$; we recover the homogeneous wave equation:
\begin{eqnarray}
    \omega^{\prime 2}\ell_{m}c(x)\left(1 + \frac{gIm\{Z_{s}(\omega)\}}{\omega^{\prime}\ell_{m}}\right)\tilde{\phi}(x,\omega^{\prime}) + \frac{\partial^{2}\tilde{\phi}(x,\omega^{\prime})}{\partial x^{2}} = 0
\end{eqnarray}
as $Re\{Z_{s}(\omega)\} = 0$ for $\hbar\omega < 2\Delta$. 
We now demonstrate that the solutions of the Euler-Lagrange equations provide the correct basis for the canonical quantization and lead to desired commutation relations between the node flux $\phi(x,t)$ and it's conjugate momenta $\rho(x,t)$, defined by Eq. $(S24)$. 
\subsection*{\large Bi-orthogonal Eigenfunctions}   
\twocolumngrid
In this section we describe the orthogonality and completeness relations satisfied by the solutions of Eqs. $(8)$ \& $(9)$ in the main text. Typically, solutions to wave-equations are expressed in terms of the eigenfunctions of the Helmholtz operator, however, 
due to the non-hermitian nature of the wave equation above the superconducting gap, orthogonal eigenfunctions do not exist. Note that the non-hermitian nature stems from the superconductor's surface resistance, hence the case here differs from the open-cavity situation typically considered in cavity-QED systems\cite{OpenCavityQED1}. Nevertheless, a set of bi-orthogonal eigenfunctions can be found by considering the eigenfunctions\cite{morse1946methods} of the operator $\mathcal{L}$ and it's adjoint $\mathcal{L}^{\dagger}$:
\begin{eqnarray}
    \mathcal{L}(\omega) &&\equiv \frac{1}{\ell_{m}c(x)}\frac{\partial^{2}}{\partial x^{2}} - i \frac{g\omega Z_{s}(\omega)}{\ell_{m}}\\ \mathcal{L}^{\dagger}(\omega) &&\equiv \frac{1}{\ell_{m}c(x)}\frac{\partial^{2}}{\partial x^{2}} + i \frac{g\omega Z^{\star}_{s}(\omega)}{\ell_{m}}
\end{eqnarray}
We consider Dirichlet boundary conditions. The eigenvalues, denoted as $\omega_{n}(\omega)$ and $\omega_{m}^{\prime}(\omega)$ implicitly depend on the frequency $\omega$ since the differential operator $\mathcal{L}$ \& $\mathcal{L}^{\dagger}$ are functions of $\omega$ and are given by:
\begin{eqnarray}
    \mathcal{L}\Psi_{n}(x,\omega_{n}) &&= -\omega_{n}^{2}\Psi_{n}(x,\omega_{n}) \\
    \mathcal{L}^{\dagger}\Phi_{m}(x,\omega^{\prime}_{m}) &&= -\omega^{\prime 2}_{m}\Phi_{m}(x,\omega^{\prime}_{m})
\end{eqnarray}
Then, given $\omega$ the sets $\{\omega_{n}(\omega)\}$ and $\{\omega_{m}^{\prime}(\omega)\}$ are complex conjugates of each other. For conciseness we drop this implicit dependence on frequency. We now prove that the sets of eigenfunctions, determined by Eqs. $(38)$ and $(39)$ are complete. Firstly, we note the following bi-orthogonality relation:
\begin{eqnarray}
\int dx \ell_{m}c(x)\Phi_{m}^{\star}(x,\omega_{m}^{\prime})\Psi_{n}(x,\omega_{n}) = \gamma(\omega)\delta_{mn}
\end{eqnarray}
where, in general, the right-hand side also depends on frequency for consistency. However, we can set $\gamma(\omega)$ to unity assuming that the functions $\Psi_{n}$ and $\Phi_{m}^{\star}$ are appropriately normalized. 
\subsection*{Completeness Relation}
We first prove that the set of eigenfunctions $\{\Psi_{n}(x,\omega_{n}(\omega))\}$ is linearly independent. Consider that the set $\{\Psi_{n}(x,\omega_{n}(\omega)\}$ is linear dependent: this implies that there exists constants $c_{n}$ such that: $\sum_{n}c_{n}\Psi_{n} = 0$. Then, by taking the inner product of the above sum with some adjoint eigen-function $\Phi_{m}(x,\omega_{m}^{\prime}(\omega))$ we get that:
\begin{align*}
\sum_{n}c_{n} \int dx w(x,\omega)\Phi_{m}^{\star}(x,\omega^{\prime}_{m}(\omega))\Psi_{n}(x,\omega_{n}(\omega)) = c_{m}
\end{align*}
which implies that $c_{m}=0$. Therefore the sets $\{\Psi_{n}\}_{n=1}^{n=\infty}$ and $\{\Phi_{m}\}_{m=1}^{m=\infty}$ are linearly independent. We now prove that the set of bi-orthogonal eigenvectors can be used to express any function, and therefore, form a bi-orthogonal basis.\\\\ Consider an abritrary function $f(x)$, and evaluate the following sum:
\begin{align}
    \sum_{n}\left(\int dx^{\prime}\ell_{m} c(x^{\prime}) \Phi_{n}^{\star}(x^{\prime},\omega_{n})f(x^{\prime})\right)\Psi_{n}(x,\omega_{n}) \equiv g(x)
\end{align}
where we assume that the above sum evaluates to a function $g(x)$. Now, the overlap of $g(x)$ with the conjugate basis $\{\Phi_{n}(x,\omega)\}$ is given by:
\begin{eqnarray}
    &&\int dx \ell_{m}c(x)\Phi_{n^{\prime}}^{\star}(x,\omega_{n}^{\prime})g(x) \nonumber \\ &&= \sum_{n}\int dx dx^{\prime}\ell_{m}c(x)\ell_{m}c(x^{\prime})\Phi_{n^{\prime}}^{\star}(x,\omega_{n^{\prime}})\Phi_{n}^{\star}(x^{\prime},\omega_{n})f(x^{\prime})\Psi_{n}(x,\omega_{n}) \nonumber\\
    &&= \sum_{n}\delta_{nn^{\prime}}\int dx^{\prime}\ell_{m}c(x^{\prime})\Phi_{n}^{\star}(x^{\prime},\omega_{n})f(x^{\prime}) \nonumber\\
    &&= \int dx^{\prime} \ell_{m}c(x^{\prime})\Phi_{n^{\prime}}^{\star}(x^{\prime},\omega_{n}^{\prime})f(x^{\prime})
\end{eqnarray}
Therefore, the overlap of $g(x)$ and $f(x)$ with the basis set $\{\Phi_{n}\}$ are simply identical. Given the linear independence of the bi-orthogonal sets, we must have that: $f(x) = g(x)$. Then, by plugging in $g(x) = f(x)$ in $(41)$, we get:
\begin{eqnarray}
\sum_{n}\ell_{m} c(x^{\prime})\Phi^{\star}_{n}(x^{\prime},\omega_{n}^{\prime})\Psi_{n}(x,\omega_{n}) = \delta(x-x^{\prime})
\end{eqnarray}
We refer to articles \cite{QNMStephen1,QNMStephen2,QNMTutorialKurt,QNMYoungPRA1,QNMYoungPRA2,RouteHAKAN} for a straightforward generalization of Eqs. $(S40)$ \& $(S43)$ for resonators with radiative boundary conditions accounting for losses due to the field escaping from the boundaries. In the context of circuit-QED systems these losses have been included directly in the quantization procedure\cite{HakanA2,NMHakan}. Finally, the Green's function in terms of the bi-orthogonal modes is given as: 
\begin{eqnarray}
     G(x,x^{\prime};\omega) = \sum_{n}\frac{\Psi_{n}(x,\omega_{n}(\omega))\Phi_{n}^{\star}(x^{\prime},\omega^{\prime}_{n}(\omega))}{\omega^{2} - \omega_{n}^{2}(\omega)}
\end{eqnarray}
\subsection*{\large Solution to the Euler-Lagrange Equations: Below superconducting-gap}
Below, the superconducting-gap $R_{s}(\omega) =0$. Then,
the solution to the homogeneous equation of motion $(S35)$ is given by: 
\begin{align}
    \tilde{\phi}(x,\omega) = \sum_{n}c_{n}\delta(\omega-\omega_{n})\tilde{\Psi}_{n}(x,\omega_{n})
\end{align}
where each of the term above satisfies equation $(35)$, therefore by linearity so does the sum. In the time domain the flux satisfies:
\begin{align*}
    \phi(x,t) &= \sum_{n}\left(c_{n}^{\star}e^{i\omega_{n}t} + c_{n}e^{-i\omega_{n}t}\right)\tilde{\Psi}_{n}(x,\omega_{n}) \\
    \rho(x,t) &= i\sum_{n}\omega_{n}\left(c_{n}^{\star}e^{i\omega_{n}t} - c_{n}e^{-i\omega_{n}t}\right)\tilde{\Psi}_{n}(x,\omega_{n})
\end{align*}
where we have used the fact that the eigenmodes are real. The sum now runs over eigenmodes with eigen-frequencies in the interval $(0, 2\Delta)$. The corresponding quantum versions are obtained by promoting $c_{n}$ to operators and imposing that they satisfy:
\begin{align*}
    [\hat{c}_{n},\hat{c}_{n}^{\dagger}] = |\alpha_{n}|^{2}
\end{align*}
where we can fix $\alpha_{n}$ from the final commutation relation between $\hat{\phi}(x)$ and $\hat{\rho}(y)$. 
\onecolumngrid
\subsection*{\large Solution of the Euler-Lagrange Equations: Modes above the superconducting-gap}
For frequencies above the superconducting-gap, we can utilize the Green's function to provide a solution for equation $(S35)$:
\begin{align}
    \tilde{\phi}(x,\omega) = \int_{0}^{L} dx^{\prime} G(x,x^{\prime};\omega)j_{N}(x^{\prime};\omega)
\end{align}
where the operator $j_{N}(x,\omega) = -i\sqrt{g\ell_{m}Re\{Z_{s}(\omega)\}c(x)}\omega \mathbf{X}_{\omega}(x)$. Therefore, above the superconducting-gap the flux is expressed as:
\begin{eqnarray}
    \phi(x) &&= -i\int_{0}^{\infty}d\omega \int_{0}^{L} dx G(x,x^{\prime};\omega)\sqrt{g\ell_{m}Re\{Z_{s}(\omega)\}c(x^{\prime})}\omega \mathbf{X}_{\omega}(x^{\prime}) \nonumber \\
    && \;\;\;\; + i\int_{0}^{\infty}d\omega \int_{0}^{L} dx G^{\star}(x,x^{\prime};\omega)\sqrt{g\ell_{m}Re\{Z_{s}(\omega)\}c(x^{\prime})}\omega \mathbf{X}^{\star}_{\omega}(x^{\prime})
\end{eqnarray}
In order to get the charge density $\rho(x,t)$ in terms of the homogeneous solution $\mathbf{X}_{\omega}(x)$, we recall that the solution $\psi_{\omega}$ satisfies: 
\begin{eqnarray}
    \psi_{\omega}(x,\omega^{\prime}) = -\sqrt{\frac{2}{\pi}}\sqrt{\frac{gRe\{Z_{s}(\omega)\}c(x)}{\ell_{m}}} \frac{(i\omega^{\prime})\tilde{\phi}(x,\omega^{\prime})}{\omega^{2} - \omega^{\prime 2}} + \delta(\omega - \omega^{\prime})\mathbf{X}_{\omega^{\prime}}(x) + \delta(\omega + \omega^{\prime})\mathbf{X}^{\star}_{\omega^{\prime}}(x) \nonumber 
\end{eqnarray}
In the above equation, we can plug in the solution for the flux $\tilde{\phi}(x,\omega)$ in terms of the Green's function which gives:
\begin{eqnarray}
    \psi_{\omega}(x,\omega^{\prime}) &&= -\sqrt{\frac{2}{\pi}}\sqrt{\frac{gRe\{Z_{s}(\omega)\}c(x)}{\ell_{m}}} \frac{\omega^{\prime 2}}{\omega^{2} - \omega^{\prime 2}}\int_{0}^{L}dx G(x,x^{\prime};\omega^{\prime})\sqrt{g\ell_{m}Re\{Z_{s}(\omega^{\prime})\}c(x^{\prime})}\mathbf{X}_{\omega^{\prime}}(x^{\prime}) \nonumber \\ && \;\;\;\; + \delta(\omega - \omega^{\prime})\mathbf{X}_{\omega^{\prime}}(x) + \delta(\omega + \omega^{\prime})\mathbf{X}^{\star}_{\omega^{\prime}}(x)
\end{eqnarray}
The conjugate momenta can be expressed in terms of it's Fourier transform as:
\begin{eqnarray}
    \tilde{\rho}(x,\omega^{\prime}) = c(x)(i\omega^{\prime})\tilde{\phi}(x,\omega^{\prime}) + \sqrt{\frac{2}{\pi}}\int_{0}^{\infty} d\omega \sqrt{\frac{gRe\{Z_{s}(\omega)\}c(x)}{\ell_{m}}}\tilde{\psi}_{\omega}(x,\omega^{\prime})
\end{eqnarray}
In the above equation, we express each of $\tilde{\phi}$ and $\tilde{\psi}$ in terms of the homogeneous solution $\mathbf{X}_{\omega}$. This gives the following terms: 
\begin{align*}
    c(x)(i\omega^{\prime})\tilde{\phi}(x,\omega^{\prime}) = c(x)(\omega^{\prime 2})\int_{0}^{L}dx^{\prime}G(x,x^{\prime};\omega^{\prime})\sqrt{g\ell_{m}Re\{Z_{s}(\omega)\}c(x)}\mathbf{X}_{\omega^{\prime}}(x^{\prime})
\end{align*}
and the second term gives:
\begin{align*}
    \frac{2}{\pi}\int_{0}^{\infty}d\omega \frac{gRe\{Z_{s}(\omega)\}c(x)}{\ell_{m}}\frac{1}{\omega^{2} - \omega^{\prime 2}}\int_{0}^{L}dx \omega^{\prime 2}G(x,x^{\prime};\omega^{\prime})\sqrt{g\ell_{m}Re\{Z_{s}(\omega^{\prime})\}c(x^{\prime})}\mathbf{X}_{\omega^{\prime}}(x^{\prime})
\end{align*}
In the above equation we can utilise Kramer-Kronig relations to complete the integrals over $\omega$, which gives us: 
\begin{align*}
&c(x)\int_{0}^{L}dx^{\prime}\omega^{\prime 2}G(x,x^{\prime};\omega^{\prime})\left(\frac{gIm\{Z_{s}(\omega^{\prime})\}}{\omega^{\prime}\ell_{m}} - i\frac{gRe\{Z_{s}(\omega^{\prime})\}}{\omega^{\prime}\ell_{m}}\right)\sqrt{g\ell_{m}Re\{Z_{s}(\omega^{\prime})\}c(x^{\prime})}\mathbf{X}_{\omega^{\prime}}(x^{\prime}) 
\end{align*}
Finally, we add the two contributions, one from the equation above and the other from $c(x)(i\omega^{\prime})\tilde{\phi}(x,\omega^{\prime})$ to give: 
\begin{align*}
    c(x)\int_{0}^{L}dx^{\prime}\omega^{\prime 2}G(x,x^{\prime};\omega^{\prime})\epsilon(\omega^{\prime})\sqrt{g\ell_{m}Re\{Z_{s}(\omega^{\prime})\}c(x^{\prime})}\mathbf{X}_{\omega^{\prime}}(x^{\prime})
\end{align*}
At last, there is also a direct contribution to $\rho$ from the solution $\mathbf{X}$, which is:
\begin{align*}
    \sqrt{\frac{2}{\pi}}\sqrt{\frac{Re\{Z_{s}(\omega)\}c(x)}{\ell_{m}}}\mathbf{X}_{\omega}(x)
\end{align*}
Therefore, the contribution of the modes \textbf{above the superconducting gap} to the solutions for the fields $\hat{\phi}(x)$ and $\hat{\rho}(x)$ read as:
\begin{eqnarray}
    \phi(x) &&= -i\int_{0}^{\infty}d\omega \int_{0}^{L} dx G(x,x^{\prime};\omega)\sqrt{g\ell_{m}Re\{Z_{s}(\omega)\}c(x^{\prime})}\omega \mathbf{X}_{\omega}(x^{\prime}) \nonumber \\
    && \;\;\;\; + i\int_{0}^{\infty}d\omega \int_{0}^{L} dx G^{\star}(x,x^{\prime};\omega)\sqrt{g\ell_{m}Re\{Z_{s}(\omega)\}c(x^{\prime})}\omega \mathbf{X}^{\star}_{\omega}(x^{\prime})
\end{eqnarray}
\begin{eqnarray}
    \rho(x) &&= c(x)\int_{0}^{\infty} d\omega \int_{0}^{L}dx^{\prime}\omega^{2}G(x,x;\omega)\epsilon(\omega)\sqrt{g\ell_{m}Re\{Z_{s}(\omega^{\prime})\}c(x^{\prime})}\mathbf{X}_{\omega^{\prime}}(x^{\prime}) \nonumber\\
    &&\;\;\;\; + c(x)\int_{0}^{\infty} d\omega \int_{0}^{L}dx^{\prime}\omega^{2}G^{\star}(x,x;\omega)\epsilon^{\star}(\omega)\sqrt{g\ell_{m}Re\{Z_{s}(\omega^{\prime})\}c(x^{\prime})}\mathbf{X}^{\star}_{\omega^{\prime}}(x^{\prime}) \nonumber \\
    && \;\;\;\; + \int_{0}^{\infty}d\omega\sqrt{\frac{2}{\pi}}\sqrt{\frac{gRe\{Z_{s}(\omega)\}c(x)}{\ell_{m}}}\mathbf{X}_{\omega}(x) + \int_{0}^{\infty}d\omega\sqrt{\frac{2}{\pi}}\sqrt{\frac{gRe\{Z_{s}(\omega)\}c(x)}{\ell_{m}}}\mathbf{X}_{\omega}^{\star}(x)
\end{eqnarray}
\subsection*{\large Green's Identity}
The Green's function satisfies certain properties which come into use for quantization, specifically to demonstrate $[\hat{\phi}(x),\hat{\rho}(y)] = \delta(x-y)$. We prove that: 
\begin{align}
    g\omega R_{s}(\omega)\int dx^{\prime}c(x^{\prime})G^{\star}(x_{1},x^{\prime};\omega)G(x,x^{\prime};\omega) = i\frac{G^{\star}(x_{1},x;\omega) - G(x,x_{1};\omega)}{2}
\end{align}
We utilize two properties of the Green's function: 1. Symmetry under exchange of $x$ and $x^{\prime}$, $G(x,x^{\prime};\omega) = G(x^{\prime},x;\omega)$, 2. Complex conjugate symmetry $G(x,x^{\prime}, -\omega) = G^{\star}(x,x^{\prime},\omega)$, which follows from the fact that: $Z^{\star}(\omega) = Z(-\omega)$.
We start off with the symmetry property: 
\begin{align*}
     \frac{\partial^{2} G(x,x^{\prime};\omega)}{\partial x^{\prime 2}} + \omega^{2}c(x^{\prime})\ell_{m}\left(1 + \frac{g}{i\omega \ell_{m}}Z_{s}(\omega)\right)G(x,x^{\prime};\omega) = \delta(x-x^{\prime})
\end{align*}
Multiplying the above equation with $G^{\star}(x_{1},x^{\prime};\omega)$ and integrating over $x^{\prime}$, gives us:
\begin{align}
    &\omega^{2}\ell_{m}\left(1 + \frac{g}{i\omega \ell_{m}}Z_{s}(\omega)\right)\int dx^{\prime} c(x^{\prime})G^{\star}(x_{1},x^{\prime};\omega)G(x,x^{\prime};\omega)\\ & \;\;\;\;\;\;\;\;- \int dx^{\prime}\frac{\partial G^{\star}(x_{1},x^{\prime};\omega)}{\partial x^{\prime}}\frac{\partial G(x,x^{\prime};\omega)}{\partial x^{\prime}} = G^{\star}(x_{1},x;\omega) \notag 
\end{align}
where we used the homogeneous Neumann boundary condition satisfied by the Green's function. To arrive at $(S52)$, we take the complex conjugate of the above equation, interchange $x$ and $x_{1}$ and subtract the resulting equation from the above equation.
\subsection*{\large Equal-time Commutation relations: $[\hat{\phi}(x),\hat{\rho}(y)]$}
The commutation relation between the conjugate fields $\hat{\phi}(x)$ and $\hat{\rho}(x)$ follow from the properties of the Green's function and the commutation relations followed by the fields $\mathbf{X}_{\omega}(x)$ and $\mathbf{X}^{\star}_{\omega}(x^{\prime})$. We propose the following: 
\begin{align*}
     [\hat{\mathbf{X}}_{\omega}(x),\hat{\mathbf{X}}_{\omega^{\prime}}^{\star}(x^{\prime})] = \alpha(\omega)\delta(x-x^{\prime})\delta(\omega  - \omega^{\prime})
\end{align*}
where $\alpha(\omega)$ is to be determined later. The commutator $[\hat{\phi}(x),\hat{\rho}(y)]$ is calculated in two parts: first is the contribution from the modes below the superconducting-gap; and the second is the contribution from the modes above the superconducting-gap.
The contribution from the modes above the superconducting-gap is: 
\begin{align*}
&-i\ell_{m}c(y)\int_{0}^{\infty}d\omega \int_{0}^{L}dx^{\prime}\omega^{3}\epsilon^{\star}(\omega)\alpha(\omega)G(x,x^{\prime};\omega)G^{\star}(y,x^{\prime};\omega)\left(gRe\{Z_{s}(\omega)\}c(x^{\prime})\right) \\
& \;\;\;\; -i\ell_{m}c(y)\int_{0}^{\infty}d\omega \int_{0}^{L}dx^{\prime}\omega^{3}\epsilon(\omega)\alpha(\omega)G(y,x^{\prime};\omega)G^{\star}(x,x^{\prime};\omega)\left(gRe\{Z_{s}(\omega)\}c(x^{\prime})\right) 
\end{align*}
We now utilize the previously derived Green's identity to give: 
\begin{align*}
    -i\ell_{m}c(y)\int_{0}^{\infty}d\omega \omega^{2}\epsilon^{\star}(\omega)\alpha(\omega)Im\{G(y,x;\omega)\}  -i\ell_{m}c(y)\int_{0}^{\infty}d\omega \omega^{2}\epsilon   (\omega)\alpha(\omega)Im\{G(y,x;\omega)\} 
\end{align*}
The imaginary part of the Green's function can be written as: 
\begin{align*}
    Im\{G(y,x;\omega)\} = -\frac{G^{\star}(y,x;\omega) - G(y,x;\omega)}{2i}
\end{align*}
We now use the fact that: $G^{\star}(y,x;\omega) \equiv G(x,y;-\omega)$, giving us:
\begin{align*}
    &\frac{\ell_{m}c(y)}{2}\int_{0}^{\infty}d\omega \omega^{2}\epsilon^{\star}(\omega)\alpha(\omega)\left(G(x,y;-\omega) - G(x,y;\omega)\right)\\ &+ \frac{\ell_{m}c(y)}{2}\int_{0}^{\infty}d\omega \omega^{2}\epsilon(\omega)\alpha(\omega)\left(G(y,x;-\omega) - G(x,y;\omega)\right)
\end{align*}
We choose $\alpha(\omega)$ to be:
\begin{align}
    \alpha(\omega) = \frac{\hbar}{\pi\omega}
\end{align}
Thereafter, the commutation relations become:
\begin{align*}
    &\frac{\ell_{m}c(y)\hbar}{2\pi}\int_{2\Delta}^{\infty}d\omega \omega\epsilon^{\star}(\omega)\left(G(x,y;-\omega) - G(x,y;\omega)\right)\\ &+ \frac{\ell_{m}c(y)\hbar}{2\pi}\int_{2\Delta}^{\infty}d\omega \omega\epsilon(\omega)\left(G(y,x;-\omega) - G(x,y;\omega)\right)
\end{align*}
Using the properties of the Green's function we can re-arrange the above as: 
\begin{align*}
    &-\frac{\ell_{m}c(y)\hbar}{2\pi}\int_{-\infty}^{-2\Delta}d\omega\omega \epsilon(\omega)G(x,y;\omega) - \frac{\ell_{m}c(y)\hbar}{2\pi}\int_{2\Delta}^{\infty}d\omega \omega \epsilon(\omega) G(x,y;\omega) \\
    &-\frac{\ell_{m}c(y)\hbar}{2\pi}\int_{-\infty}^{-2\Delta}d\omega \omega \epsilon^{\star}(\omega) G(y,x;\omega) -\frac{\ell_{m}c(y)\hbar}{2\pi}\int_{2\Delta}^{\infty} d\omega \omega \epsilon(\omega) G(x,y;\omega) \\
    &= -\frac{\ell_{m}c(y)\hbar}{\pi}\int_{-\infty}^{-2\Delta}d\omega \omega Re\{\epsilon(\omega)\}G(x,y;\omega) - \frac{\ell_{m}c(y)\hbar}{\pi}\int_{2\Delta}^{\infty}d\omega \omega Re\{\epsilon(\omega)\} G(x,y;\omega)
\end{align*}
The above integral, can be computed using the method of residues, if the integration domain was from $(-\infty,\infty)$, however we have that domain misses out on the region $(-2\Delta, 2\Delta)$. This can be easily fixed, since the Green's function is purely real (hence even function) for the frequency interval $[-2\Delta,2\Delta]$.\\\\
Therefore, the principal value integral:
\begin{align*}
 \mathcal{P}\int_{-2\Delta}^{2\Delta}d\omega \omega Re\{\epsilon(\omega)\}G(x,y,\omega) = 0
\end{align*}
since the Green's function and the refractive index $\epsilon(\omega)$ is an even function of frequency $\omega$ in the region $(-2\Delta, 2\Delta)$. \\\\
We also have the following additional contribution to the commutation relation arising from the last terms in the expression for $\hat{\rho}(x)$ in the main text:
\begin{align*}
    &-i\sqrt{\frac{2}{\pi}}\frac{\hbar}{\pi}\int_{0}^{\infty}d\omega gRe\{Z_{s}(\omega)\}c(y)G(x,y;\omega) -i\sqrt{\frac{2}{\pi}}\frac{\hbar}{2}\int_{0}^{\infty}d\omega gRe\{Z_{s}(\omega)\}c(y)G^{\star} (x,y;\omega) \\
    &= -i\sqrt{\frac{2}{\pi}}\frac{\ell_{m}c(y)\hbar}{\pi}\int_{-\infty}^{\infty}d\omega \frac{gRe\{Z_{s}(\omega)\}}{\omega\ell_{m}}\omega G(x,y;\omega)
\end{align*}
Therefore we can write the contribution to commutation relation from the modes above the superconducting-gap as:
\begin{align}
    -\frac{\ell_{m}c(y)\hbar}{\pi}\int_{-\infty}^{\infty} d\omega \omega G(x,y;\omega)\left(1 + \frac{gIm\{Z_{s}\}}{\omega \ell_{m}} + i\sqrt{\frac{2}{\pi}}\frac{gRe\{Z_{s}(\omega)\}}{\omega \ell_{m}}\right)
\end{align}
which can be computed using the residue theorem. We choose to work with the lower half plane since the Green's function has no poles in this domain and hence we can calculate it's integral along an semi-circular arc of infinite radius. Applying the residue theorem gives:
\begin{align*}
   \int_{-\infty}^{\infty} d\omega \omega G(x,y;\omega) + \int_{C}d\omega \omega G(x,y;\omega) = \text{sum of all residues}
\end{align*}
where $C$ corresponds to the a semi-circle of infinitely large radius in the lower half complex plane. We note that the Green's function has poles on the real axis as well as poles in the upper half complex plane. The poles on the real axis lie in the frequency interval $(-2\Delta,2\Delta)$. These poles on the real axis contribute with a factor of $-i\pi$, instead of $-2\pi i$. This can be seen by deforming the contour of integration into a semi-circular arc of infinitesimal radius below the poles on the real axis. The integration over the contour $C$ is:
\begin{align*}
   \int_{C}\omega G(x,y,\omega) &=  \sum_{n}\lim_{R\rightarrow \infty}\int_{0}^{-\pi}d\theta (iRe^{i\theta})(Re^{i\theta})\frac{\Psi_{n}(x;\omega)\Phi_{n}^{\star}(y;\omega)}{R^{2}e^{2i\theta} - \omega_{n}^{2}(\omega)} \\ &= \sum_{n}\lim_{R\rightarrow \infty}\int_{0}^{-\pi}d\theta (iRe^{i\theta})(Re^{i\theta})\frac{\Psi_{n}(x;\omega)\Phi_{n}^{\star}(y;\omega)}{R^{2}e^{2i\theta} - \omega_{n}^{2}(Re^{i\theta})}
\end{align*}
Then, in the limit of $R\rightarrow \infty$, the right hand side gives:
\begin{align*}
    \lim_{R\rightarrow \infty} i \frac{R^{2}e^{2i\theta}}{R^{2}e^{2i\theta}}\frac{\Psi_{n}(x;\omega_{n}(Re^{i\theta}))\Phi_{n}^{\star}(y;\omega_{n}(Re^{i\theta}))}{1 - \frac{\omega_{n}^{2}(Re^{i\theta})}{R^{2}e^{2i\theta}}} = i\lim_{R\rightarrow \infty}\Psi_{n}(x;\omega_{n})\Phi_{n}^{\star}(y;\omega_{n})
\end{align*} 
We finally get the result that:
\begin{align*}
    \int_{C}d\omega \omega G(x,y;\omega) = -i\pi\sum_{n}\Psi_{n}(x;\omega_{n})\Phi_{n}^{\star}(y;\omega_{n}) = -\frac{i\pi}{\ell_{m}c(y)}\delta(x-y)
\end{align*}
The residue of the integrand at a pole $\omega_{n}$ (which is real and lies between $-2\Delta$ and $2\Delta$) is:
\begin{align*}
    Res_{\omega \rightarrow \omega_{n}}\left\{\omega \left(1 + \frac{gZ_{s}(\omega)}{\omega\ell_{m}}\right) G(x,y,\omega)\right\}  = \frac{1}{2}\left(1 + \frac{gZ_{s}(\omega_{n})}{\omega_{n}\ell_{m}}\right)\Psi_{n}(x;\omega_{n})\Phi^{\star}_{n}(y;\omega_{n})
\end{align*}
Finally we get that:
\begin{align}
   \mathcal{P}\int_{-\infty}^{\infty}d\omega \omega G(x,y,\omega) = \frac{-\pi i}{2}\left(1 + \frac{gZ_{s}(\omega_{n})}{\omega_{n}\ell_{m}}\right)\sum_{|Re(\omega_{n})| < 2\Delta}\Psi_{n}(x;\omega_{n})\Phi^{\star}_{n}(y;\omega_{n}) + \frac{i\pi}{\ell_{m}c(y)}\delta(x-y)
\end{align}
Therefore, the modes above the superconducting gap give: 
\begin{align*}
    \frac{-i\hbar}{2}\ell_{m}c(y)\left(1 + \frac{gZ_{s}(\omega_{n})}{\omega_{n}\ell_{m}}\right)\sum_{|Re(\omega_{n})| < 2\Delta}\Psi_{n}(x;\omega_{n})\Phi^{\star}_{n}(y;\omega_{n}) + i\hbar\delta(x-y)
\end{align*}
The contribution from the modes below the superconducting gap is:
\begin{align*}
i\sum_{0< \omega_{n} < 2\Delta}2|\alpha_{n}|^{2}\omega_{n}\Psi_{n}(x,\omega_{n})\Psi_{n}(y,\omega_{n})
\end{align*}
This implies that we can choose $\alpha_{n}$ to be:
\begin{align*}
    \alpha_{n} = \sqrt{\frac{\hbar\ell_{m}}{2\omega_{n}}}\sqrt{\epsilon(\omega_{n})}
\end{align*}
Finally, adding the contribution from the modes below the superconducting-gap gives the commutation relation:
\begin{align}
    \boxed{[\hat{\phi}(x),\hat{\rho}(y)] = i\hbar\delta(x-y)}
\end{align}
    \subsection*{\large Extension of Mattis Bardeen Conductivity to Complex Domain}
In the following sections we describe the analysis required to extend the conductivity to complex frequencies. We describe a standard procedure, closely following the analysis by Mattis \& Bardeen. Finally, we also describe the transformations required to convert the integral expressions provided in the main text to integrals of the elliptic type, which can be evaluated exactly. \\\\
Mattis and Bardeen consider the effect of an external magnetic field on the superconductor by including it's effect perturbatively. They consider an expansion of the Cooper pair wave function to the first order in the field strength, effectively considering linear response theory. Then, arbitrary fields can be simply considered by superposition. An extension of the conductivity to complex frequencies is straightforward by considering a complex addition to the frequency $\omega$ in the expression for $L(\omega,\epsilon,\epsilon^{\prime})$ in \cite{mattis1958theory}, as suggested by Mattis \& Bardeen in the context of including the effects of electron scattering. The conductivity is derived by the following relation: 
\begin{align}
    \frac{\sigma(\omega, \kappa)}{\sigma_{n}} = \frac{I(\omega, R, T)}{-2\pi i(\omega-i\kappa)}
\end{align}
where the integral $I(\omega, R, T)$, known as the Mattis-Bardeen kernel\cite{mattis1958theory}, is given by:
\begin{align}
    I(\omega, R, T) = \int_{-\infty}^{\infty}\int_{-\infty}^{\infty} d\epsilon d\epsilon^{\prime} L(\omega, \epsilon, \epsilon^{\prime})\cos(\alpha (\epsilon-\epsilon^{\prime}))
\end{align}
where $\alpha = \frac{R}{\hbar v_{0}}$ \& $v_{o}$ the Fermi velocity. The kernel $I(\omega,\epsilon,\epsilon^{\prime})$ depends on the frequency and the temperature through the expression $L(\omega, \epsilon, \epsilon^{\prime})$. The zero temperature expression for $L(\omega, \epsilon,\epsilon^{\prime})$ is\footnote{The following expression for $L(\omega,\epsilon,\epsilon^{\prime})$ differs from the one in the main text in the sense that $\kappa \rightarrow -\kappa$. We note that this is a simple change in convention of defining the Surface Impedance as either $Z \sim (\sigma_{1} + i\sigma_{2})^{\gamma}$ or $Z \sim (\sigma_{1} - i\sigma_{2})^{\gamma}$. The plus sign implies that $Z(\omega)$ is analytic in the upper half complex plane.}: 
\begin{align}
    L(\omega, \epsilon, \epsilon^{\prime}) = \left(1 - \frac{\epsilon\epsilon^{\prime} + \Delta^{2}}{E E^{\prime}}\right)\left(\frac{1}{E + E^{\prime} - \hbar(\omega - i\kappa)} + \frac{1}{E + E^{\prime} + \hbar(\omega - i\kappa)}\right)
\end{align}
where $E = \sqrt{\Delta^{2} + \epsilon^{2}}$ and $E^{\prime} = \sqrt{\Delta^{2} + \epsilon^{\prime 2}}$. We evaluate the Mattis-Bardeen kernel for non-zero $\kappa$ by means of contour integration. We proceed ahead by integrating over $\epsilon^{\prime}$ first, where the appropriate choice of contour is to be made in order to simplify the calculations. We work with the principal branch of the square root, $\sqrt{z} = \sqrt{r}e^{\frac{i\theta}{2}}$, where $\theta \in (-\pi, \pi)$. The branch cut lies along the negative real axis for $\sqrt{z}$. Further, we note the convention that for real $x$, $\sqrt{x^{2}} = \pm |x|$, with the negative sign arising for the negative Riemann sheet of the square root. \\\\
Before we proceed ahead, we note that the location of these poles does not depend on the temperature. Therefore, we only consider the $T=0$ case where closed form expressions can be derived. The extension to non-zero temperature case is straightforward, however, closed form expressions can only be obtained under considerable approximations\cite{gao2008physics}.\\\\
The branch cut of $E^{\prime} = \sqrt{\Delta^{2} + \epsilon^{\prime 2}}$ lies on the imaginary axis, between $i\Delta$ and $\infty$, as well as between $-\infty$ and $-i\Delta$. The branch cut of $L(\omega, \epsilon, \epsilon^{\prime})$, as a function of $\epsilon^{\prime}$, is also along the same interval. In order to proceed with an appropriate contour, we need the poles of $L(\omega,\epsilon,\epsilon^{\prime})$. The poles, which may be complex, are identified by setting:
\begin{align}
    \epsilon^{\prime 2} + \Delta^{2} = re^{i\theta}  \;\;\;\;\;\; \theta \in [-\pi, \pi]
\end{align}
and are given by the solutions of:
\begin{align*}
    \sqrt{r}e^{\frac{i\theta}{2}} + \sqrt{\epsilon^{2} + \Delta^{2}} + \hbar(\omega - i\kappa) &= 0  \tag{A1}\\
    \sqrt{r}e^{\frac{i\theta}{2}} + \sqrt{\epsilon^{2} + \Delta^{2}} -\hbar(\omega - i\kappa) &= 0 \tag{A2}
\end{align*}
We note that the equations (A1) and (A2) will give poles lying on the negative Riemann Sheet when $E \pm \hbar\omega$ is positive. \\
\subsubsection*{\large Poles from equation (A1)}
The real and imaginary parts satisfy:
\begin{align*}
    \sqrt{r}\cos\left(\frac{\theta}{2}\right) + \sqrt{\epsilon^{2} + \Delta^{2}} + \hbar\omega &= 0 \\
    \sqrt{r}\sin\left(\frac{\theta}{2}\right) - \hbar\kappa &= 0 
\end{align*}
These equations can be solved to give $\epsilon^{\prime}$: 
\begin{align}
    \epsilon^{\prime 2} &= r\left(\cos^{2}\frac{\theta}{2} - \sin^{2}\frac{\theta}{2}\right) - \Delta^{2} + 2ir\sin\frac{\theta}{2}\cos\frac{\theta}{2} \notag \\
    &= \left(\sqrt{\epsilon^{2} + \Delta^{2}} + \hbar\omega \right)^{2} - (\hbar\kappa)^{2} -\Delta^{2} - 2i(\hbar\kappa)(\sqrt{\epsilon^{2} + \Delta^{2}} + \hbar\omega)
\end{align}
The limit $\kappa \rightarrow 0^{+}$ should reconcile our solution with results from Mattis-Bardeen theory. Firstly, note that we must work with the negative value of the square root: As $\kappa \rightarrow 0^{+}$, we have that $\epsilon^{\prime 2} + \Delta^{2} \rightarrow \left(\hbar\omega + \sqrt{\epsilon^{2} + \Delta^{2}}\right)^{2}$, therefore implying that we should take the negative root in order to satisfy A1. Therefore, for non-zero $\kappa$, the pole lies on the \textbf{negative Riemann Sheet}. 
\begin{align}
    \epsilon^{\prime} &= -\sqrt{\left(\sqrt{\epsilon^{2} + \Delta^{2}} + \hbar\omega \right)^{2} - (\hbar\kappa)^{2} -\Delta^{2} - 2i(\hbar\kappa)(\sqrt{\epsilon^{2} + \Delta^{2}} + \hbar\omega)} \notag \\
    &= -\sqrt{\left(E + \hbar(\omega - i\kappa) \right)^{2} - \Delta^{2}}
\end{align}
As $\hbar\omega + E \geq \Delta$, we have that the pole lies in the \textbf{second quadrant} (specifically, between $\frac{3\pi}{4}$ and $\pi$). As $\kappa \rightarrow 0^{+}$, the pole tends to the negative real axis, giving: 
\begin{align}
    \epsilon^{\prime} = -\sqrt{(\sqrt{\epsilon^{2} + \Delta^{2}} + \hbar\omega)^{2} - \Delta^{2}}
\end{align}
\begin{center}
    \subsubsection*{\large Poles from Equation (A2)}
\end{center}
The real and imaginary parts satisfy: 
\begin{align}
    \sqrt{r}\cos\left(\frac{\theta}{2}\right)   &= \hbar\omega - \sqrt{\epsilon^{2} + \Delta^{2}} \\
    \sqrt{r}\sin\left(\frac{\theta}{2}\right)  &= -\hbar\kappa
\end{align} 
\textbf{Case I - $\hbar\omega > \sqrt{\epsilon^{2} + \Delta^{2}}$}: We note that the above equation is satisfied for some $\frac{\theta}{2}$ in the fourth quadrant, leading to a value of $\theta \in [-\pi,0]$. As done previously, these equations can be easily inverted to give:  
\begin{align}
    \epsilon^{\prime 2} 
   &= (\hbar\omega - E - \Delta)(\hbar\omega - E + \Delta) - (\hbar\kappa)^{2} - 2i(\hbar\kappa)(\hbar\omega - E) \implies \epsilon^{\prime} = \sqrt{(\hbar(\omega - i\kappa) - E)^{2} - \Delta^{2}}
\end{align}
Note that $\kappa \rightarrow 0^{+} \implies \epsilon^{\prime 2} + \Delta^{2} \rightarrow (\hbar\omega - E)^{2}$. Thus, equation A2 is satisfied when the pole lies on the \textbf{positive Riemann Sheet}. We note two additional sub-cases: $\hbar\omega - E > \Delta$, and $\hbar\omega - E < \Delta$. \\
\begin{itemize}
    \item For $\hbar\omega - E> \Delta$, we have that $\epsilon^{\prime 2}$ (by simple extension $\epsilon^{\prime}$ as well) lies in the \textbf{fourth quadrant}. In the limit of $\kappa \rightarrow 0^{+}$, we have that $\epsilon^{\prime} \rightarrow \sqrt{(\hbar\omega - E - \Delta)(\hbar\omega - E + \Delta)}$, which is on the real axis. As $E \geq \Delta$, the condition $\hbar\omega - E >\Delta$ only holds if $\hbar\omega \geq 2\Delta$. 
    \item For $\hbar\omega - E < \Delta$, we have that $\epsilon^{\prime 2}$ lies in the \textbf{third quadrant} (hence $\epsilon^{\prime}$ belongs to the \textbf{fourth quadrant}). In the limit of $\kappa \rightarrow 0^{+}$, $\epsilon^{\prime} \rightarrow -i\sqrt{(E + \Delta - \hbar\omega)(\hbar\omega - E + \Delta)}$, since the branch cut for $\sqrt{z}$ runs along the negative real axis. This condition can be satisfied for either $\hbar\omega > 2\Delta$ or $\hbar\omega \leq 2\Delta$. 
\end{itemize}
\textbf{Case II - $\hbar\omega < \sqrt{\epsilon^{2} + \Delta^{2}}$}: is analysed similarly by considering the real and imaginary parts: 
\begin{align*}
    \sqrt{r}\cos\left(\frac{\theta}{2}\right)   &= \hbar\omega - \sqrt{\epsilon^{2} + \Delta^{2}} \\
    \sqrt{r}\sin\left(\frac{\theta}{2}\right)  &= -\hbar\kappa
\end{align*}
As before, we solve for $\epsilon^{\prime}$ by:
\begin{align}
  \epsilon^{\prime 2} 
   &= (\hbar\omega - E - \Delta)(\hbar\omega - E + \Delta) - (\hbar\kappa)^{2} - 2i(\hbar\kappa)(\hbar\omega - E)  
\end{align}
For $\kappa \rightarrow 0^{+}$, $\epsilon^{\prime 2} + \Delta^{2} \rightarrow (\hbar\omega - E)^{2}$, implying that the pole belongs to the \textbf{negative Riemann sheet}. Like before, we note that the cases $\hbar\omega - E > -\Delta$ and $\hbar\omega - E < -\Delta$ differ in the limit $\kappa \rightarrow 0^{+}$. 
\begin{align}
    \epsilon^{\prime} &= -\sqrt{(\hbar\omega - E - \Delta)(\hbar\omega - E + \Delta) - (\hbar\kappa)^{2} - 2i(\hbar\kappa)(\hbar\omega - E) }  \notag \\
    &= -\sqrt{(\hbar(\omega - i\kappa) - E)^{2} - \Delta^{2}}
\end{align}
\begin{itemize}
    \item For $\hbar\omega - E > -\Delta$, we have that $\epsilon^{\prime}$ belongs to the \textbf{third quadrant}, and in the limit $\kappa\rightarrow 0^{+}$ tends to $-i\sqrt{(E + \Delta - \hbar\omega)(\hbar\omega - E + \Delta)}$, which holds regardless of $\hbar \omega$ \& $2\Delta$. 
    \item For $\hbar\omega - E< -\Delta$, we have that $\epsilon^{\prime}$ belongs to the \textbf{third quadrant}, and in the limit $\kappa \rightarrow 0^{+}$ tends to $-\sqrt{(\hbar\omega - E - \Delta)(\hbar\omega - E + \Delta)}$.  
\end{itemize}
\begin{center}
    \subsection*{\large Contour Integration}
\end{center}
The correct choice of contour is made by considering which half of the complex plane the integrand in $(2)$ converges in. Note that $
   \cos(\alpha(\epsilon-\epsilon^{\prime})) = \frac{e^{i\alpha(\epsilon-\epsilon^{\prime})} + e^{-i\alpha(\epsilon-\epsilon^{\prime})}}{2}$
thus,  Jordan's lemma makes the choice of contour clear for each of the integrands. Further, we note that there is a contribution from the branch-cut discontinuities, which totals out to zero by an appropriate choice of contour.
In order to evaluate the integrals, we need the residue at the poles found above. These residues can be found by a Laurent Series expansion of the integrand. Consider the Laurent series of the functions $f_{\pm}(z)$ about $z_{0}$, where $z_{0}$ satisfies $\sqrt{z_{0}^{2} + \Delta^{2}} \pm a = 0$:  
\begin{align*}
    f_{\pm}(z) = \frac{1}{\sqrt{z^2 + \Delta^{2}} \pm a} = \frac{\frac{\sqrt{z_{0}^{2} + \Delta^{2}}}{z_{0}}}{z-z_{0}} - \frac{\Delta^{2}}{2z_{0}^{2}} + O(z-z_{0})
\end{align*}
We note that the poles are simple poles, and the residue is simply given by $\frac{\sqrt{z_{0}^{2} + \Delta^{2}}}{z_{0}}$.\\\\
Therefore, given the location of the poles, we note that the total contribution comes from the following integrals $(I = I_{1} + I_{2})$:
\begin{align}
    I_{1} &= \frac{1}{2}\int_{-\infty}^{\infty} d\epsilon^{\prime} \left(1 - \frac{\epsilon\epsilon^{\prime} + \Delta^{2}}{E E^{\prime}}\right)\left(\frac{1}{E + E^{\prime} + \hbar(\omega - i\kappa)}\right)e^{-i\alpha(\epsilon-\epsilon^{\prime})} \\
     I_{2} &= \frac{1}{2}\int_{-\infty}^{\infty} d\epsilon^{\prime} \left(1 - \frac{\epsilon\epsilon^{\prime} + \Delta^{2}}{E E^{\prime}}\right)\left(\frac{1}{E + E^{\prime} - \hbar(\omega - i\kappa)}\right)e^{i\alpha(\epsilon-\epsilon^{\prime})}
\end{align}
where $I_{1}$ is closed in the upper half complex plane and $I_{2}$ is closed in the lower half complex plane. We begin by computing $I_{1}$. The only pole, from Equation (A1) lies in the upper half plane. Hence, closing in the upper half contour and applying the residue theorem gives us:
\begin{align*}
    I_{1} &= \pi i \left(1 + \frac{\Delta^{2}}{E}\frac{1}{\sqrt{\epsilon^{2} + \Delta^{2}} + \hbar(\omega - i\kappa)} - \frac{\epsilon}{E}\frac{\sqrt{(\sqrt{\epsilon^{2} + \Delta^{2}} + \hbar(\omega - i\kappa))^{2} - \Delta^{2}}}{{\sqrt{\epsilon^{2} + \Delta^{2}} + \hbar(\omega - i\kappa)}}\right) \\
    &\;\;\;\; \times \left(\frac{{\sqrt{\epsilon^{2} + \Delta^{2}} + \hbar(\omega - i\kappa)}}{\sqrt{(\sqrt{\epsilon^{2} + \Delta^{2}} + \hbar(\omega - i\kappa))^{2} - \Delta^{2}}}\right)e^{-i\alpha(\epsilon + \sqrt{(\sqrt{\epsilon^{2} + \Delta^{2}} + \hbar(\omega - i\kappa))^{2} - \Delta^{2}})}
\end{align*}
We proceed ahead with the short notation $E = \sqrt{\epsilon^{2} + \Delta^{2}}$: 
\begin{align*}
    I_{1} = \pi i \int_{-\infty}^{\infty}d\epsilon\left(\frac{E(E + \hbar(\omega - i\kappa)) + \Delta^{2}}{E\sqrt{(E + \hbar(\omega - i\kappa))^{2}-\Delta^{2}}} - \frac{\epsilon}{E}\right)e^{-i\alpha(\epsilon + \sqrt{(E + \hbar(\omega - i\kappa))^{2} - \Delta^{2}})}
\end{align*}
We proceed by splitting the integral over $\epsilon$ into two integrals over the regions $(0, \infty) $ and $ (-\infty, 0)$, and then undertaking the transformation $\epsilon \rightarrow -\epsilon$ in the second integral:
\begin{align*}
    \frac{I_{1}}{\pi i} &= 2\int_{0}^{\infty}d\epsilon \left(\frac{E(E + \hbar(\omega - i\kappa)) + \Delta^{2}}{E\sqrt{(E + \hbar(\omega - i\kappa))^{2}-\Delta^{2}}}\cos(\alpha \epsilon)\right)e^{-i\alpha(\sqrt{(E + \hbar(\omega - i\kappa))^{2} - \Delta^{2}})} \\
    & + 2i\int_{0}^{\infty}d\epsilon \frac{\epsilon}{E}\sin(\alpha \epsilon)e^{-i\alpha( \sqrt{(E + \hbar(\omega - i\kappa))^{2} - \Delta^{2}})}
\end{align*}
Finally, we make the transformation: $\epsilon \rightarrow \sqrt{E^{2} - \Delta^{2}}$, giving: 
\begin{align}
   &2 \int_{\Delta}^{\infty}dE \left(\frac{E^{2} + E\hbar(\omega - i\kappa) + \Delta^{2}}{\sqrt{E^{2} - \Delta^{2}}\sqrt{( E + \hbar(\omega - i\kappa))^{2} - \Delta^{2}}}\cos(\alpha\sqrt{E^{2} - \Delta^{2}})\right)e^{-i\alpha( \sqrt{(E + \hbar(\omega - i\kappa))^{2} - \Delta^{2}})} \notag \\
   &\;\;\;\; + 2i\int_{\Delta}^{\infty}dE\sin(\alpha \sqrt{E^{2} - \Delta^{2}})e^{-i\alpha( \sqrt{(E + \hbar(\omega - i\kappa))^{2} - \Delta^{2}})}
\end{align}
Similar to Mattis \& Bardeen, we define the following complex quantities:
\begin{align}
    \epsilon_{1} = \sqrt{E^{2} -\Delta^{2}} \;\;\;\;\;\;\;\; \tilde{\epsilon}_{2} = \sqrt{(E + \hbar(\omega - i\kappa))^{2} - \Delta^{2}} \;\;\;\;\;\;\;\;
    \tilde{g}(E) = \frac{E^{2} + E\hbar(\omega - i\kappa) + \Delta^{2}}{\epsilon_{1}\tilde{\epsilon}_{2}}
\end{align}
Using the above short notation, we get that the contribution of integral $I_{1}$ is:
\begin{align}
    \boxed{(2\pi i)\int_{\Delta}^{\infty}\left(\tilde{g}(E)\cos(\alpha\tilde{\epsilon}_{1}) + i\sin(\alpha\tilde{\epsilon}_{1})\right)e^{-i\alpha\tilde{\epsilon}_{2}}}
\end{align}
In the limit of $\kappa \rightarrow 0^{+}$ we recover the last integral in the Mattis-Bardeen result. \\\\
We turn to integral $I_{2}$ now, which must be done piecewise since the value of the poles, and the limit of $\kappa \rightarrow 0^{+}$ is sensitive to whether $\hbar\omega > 2\Delta$ or $\hbar\omega < 2\Delta$. \\
\subsection*{Beyond superconducting-gap frequencies: $\hbar\omega > 2\Delta$}
For $\hbar\omega > 2\Delta$, we need to undertake piece-wise integration in three regions, since the poles from Equation (A2), and the limit $\kappa \rightarrow 0^{+}$ depend on whether $\hbar \omega > E$ or $\hbar \omega < E$:
\begin{itemize}
    \item \textbf{Region I:} $E < \hbar\omega - \Delta \implies |\epsilon| < \sqrt{\hbar\omega (\hbar\omega - 2\Delta)}$. We denote $\epsilon_{c} = \sqrt{\hbar\omega (\hbar\omega - 2\Delta)}$. 
    \item \textbf{Region II:} $\hbar\omega - \Delta < E < \hbar\omega \implies \epsilon \in (\epsilon_{c}, \sqrt{(\hbar\omega)^{2} -\Delta^{2}}) \cup (-\sqrt{(\hbar\omega)^{2} - \Delta^{2}}, -\epsilon_{c})$. 
    \item \textbf{Region III:} $ E > \hbar\omega \implies |\epsilon| > \sqrt{(\hbar\omega)^{2} - \Delta^{2}} $. 
\end{itemize}
\textbf{For the interval $E < \hbar\omega - \Delta$}, the pole lies on the \textbf{positive Riemann sheet}. Closing in the lower half plane gives:
   \begin{align*}
&\int_{|\epsilon| < \epsilon_{c}}d\epsilon \frac{1}{2}\int_{-\infty}^{\infty} d\epsilon^{\prime} \left(1 - \frac{\epsilon\epsilon^{\prime} + \Delta^{2}}{E E^{\prime}}\right)\left(\frac{1}{E + E^{\prime} - \hbar(\omega - i\kappa)}\right)e^{i\alpha(\epsilon-\epsilon^{\prime})} \\
& = (-\pi i)\int_{|\epsilon| < \epsilon_{c}}d\epsilon \left(1 - \frac{\epsilon}{E}\frac{\sqrt{(\hbar(\omega - i\kappa) - E)^{2} - \Delta^{2}}}{\hbar(\omega - i\kappa) - E} - \frac{\Delta^{2}}{E}\frac{1}{\hbar(\omega - i\kappa) - E}\right)\\
&\;\;\;\;\;\;\;\;\; \times \frac{\hbar(\omega - i\kappa) - E}{\sqrt{(\hbar(\omega - i\kappa) - E)^{2}-\Delta^{2}}}e^{i\alpha(\epsilon - \sqrt{(\hbar(\omega - i\kappa)-E)^{2} - \Delta^{2}})} \\
&= (-2\pi i)\int_{0}^{\epsilon_{c}}d\epsilon \left(\frac{E(\hbar(\omega - i\kappa) - E) - \Delta^{2}}{E\sqrt{(\hbar(\omega - i\kappa) - E)^{2}-\Delta^{2}}}\cos(\alpha \epsilon)\right)e^{-i\alpha\sqrt{(\hbar(\omega - i\kappa)-E)^{2} - \Delta^{2}}}\\
&\;\;\;\; + (-2\pi i)\int_{0}^{\epsilon_{c}}d\epsilon \left(\frac{-i\epsilon}{E}\right)\sin(\alpha \epsilon)e^{-i\alpha\sqrt{(\hbar(\omega - i\kappa)-E)^{2} - \Delta^{2}}}
\end{align*}
We now change variables from $\epsilon$ to $\epsilon = \sqrt{E^{2} - \Delta^{2}}$, which gives:
\begin{align}
&(-2\pi i)\int_{\Delta}^{\hbar\omega - \Delta}dE\left(\frac{E(\hbar(\omega - i\kappa) - E) - \Delta^{2}}{\sqrt{E^{2} - \Delta^{2}}\sqrt{(\hbar(\omega - i\kappa) - E)^{2}-\Delta^{2}}}\cos(\alpha\sqrt{E^{2} - \Delta^{2}})\right)e^{-i\alpha\sqrt{(\hbar(\omega - i\kappa)-E)^{2} - \Delta^{2}}} \notag\\
&\;\;\;\; + (-2\pi i)\int_{\Delta}^{\hbar\omega -\Delta}dE \left(-i\right)\sin(\alpha \sqrt{E^{2} - \Delta^{2}})e^{-i\alpha\sqrt{(\hbar(\omega - i\kappa)-E)^{2} - \Delta^{2}}} \notag 
\end{align}
Finally, we undertake the transformation $E \rightarrow E^{\prime} + \hbar\omega$, giving:
\begin{align}
    (-2\pi i)&\int_{\Delta - \hbar\omega}^{-\Delta}dE^{\prime}\left(-\frac{(E^{\prime} + \hbar\omega)(E^{\prime} + i\kappa)+ \Delta^{2}}{\sqrt{(E^{\prime} + \hbar\omega) - \Delta^{2}}\sqrt{(E^{\prime} + i\kappa)^{2}-\Delta^{2}}}\cos(\alpha\sqrt{(E^{\prime} + \hbar\omega)^{2} - \Delta^{2}})\right) \notag \\ & \;\;\;\;\;\;\;\;\;\;\;\;\;\; \times e^{-i\alpha\sqrt{(E^{\prime} + i\hbar\kappa)^{2} - \Delta^{2}}} \notag \\
    & +  (-2\pi i)\int_{\Delta - \hbar\omega}^{-\Delta}dE^{\prime} (-i)\sin(\alpha\sqrt{(E^{\prime} + \hbar\omega)^{2} - \Delta^{2}})e^{-i\alpha\sqrt{(E^{\prime} + i\hbar\kappa)^{2} - \Delta^{2}}} 
\end{align}
Since $E^{\prime 2} - \Delta^{2} > 0$ and $E^{\prime} < 0$, we have that the limit $\kappa \rightarrow 0^{+} \implies \sqrt{(E^{\prime} + i\kappa)^{2} - \Delta^{2}}\rightarrow \sqrt{E^{\prime 2}-\Delta^{2}}$.
\\\\    
\textbf{For the interval $\hbar\omega - \Delta < E < \hbar\omega$}, the pole lies on the \textbf{positive Riemann sheet}. We directly state the result here:
\begin{align}
    (-2\pi i)&\int_{-\Delta}^{0}dE^{\prime}\left(-\frac{(E^{\prime} + \hbar\omega)(E^{\prime} + i\hbar\kappa)+ \Delta^{2}}{\sqrt{(E^{\prime} + \hbar\omega) - \Delta^{2}}\sqrt{(E^{\prime} + i\kappa)^{2}-\Delta^{2}}}\cos(\alpha\sqrt{(E^{\prime} + \hbar\omega)^{2} - \Delta^{2}})\right) \notag \\ & \;\;\;\;\;\;\;\;\;\;\;\;\;\; \times e^{-i\alpha\sqrt{(E^{\prime} + i\hbar\kappa)^{2} - \Delta^{2}}} \notag \\
    & +  (-2\pi i)\int_{-\Delta}^{0}dE^{\prime} (-i)\sin(\alpha\sqrt{(E^{\prime} + \hbar\omega)^{2} - \Delta^{2}})e^{-i\alpha\sqrt{(E^{\prime} + i\hbar\kappa)^{2} - \Delta^{2}}} 
\end{align}
\textbf{For the interval $\hbar\omega < E$}, the pole exists on the \textbf{negative Riemann sheet}. This gives the following result:
\begin{align}
    &(-2\pi i)\int_{0}^{\Delta}dE^{\prime}\left(\frac{(E^{\prime}+ \hbar\omega)(E^{\prime} + i\kappa) +\Delta^{2}}{\sqrt{(E^{\prime}+\hbar\omega)^{2} - \Delta^{2}}\sqrt{(E^{\prime}+i\kappa)^{2}-\Delta^{2}}}\cos(\alpha\sqrt{(E^{\prime} + \hbar\omega)^{2} - \Delta^{2}})\right) \notag \\
    &\;\;\;\;\;\;\;\;\;\;\;\;\;e^{i\alpha\sqrt{(E^{\prime}+i\kappa)^{2} -\Delta^{2}}} \notag \\
    & + (-2\pi i)\int_{0}^{\Delta}dE^{\prime}(-i)\sin(\alpha\sqrt{(E^{\prime} + \hbar\omega)^{2} - \Delta^{2}})e^{i\alpha\sqrt{(E^{\prime} + i\kappa)^{2} -\Delta^{2}}} \\
    &(-2\pi i)\int_{\Delta}^{\infty}dE^{\prime}\left(\frac{(E^{\prime}+ \hbar\omega)(E^{\prime} + i\kappa) +\Delta^{2}}{\sqrt{(E^{\prime}+\hbar\omega)^{2} - \Delta^{2}}\sqrt{(E^{\prime}+i\kappa)^{2}-\Delta^{2}}}\cos(\alpha\sqrt{(E^{\prime} + \hbar\omega)^{2} - \Delta^{2}})\right) \notag \\
    &\;\;\;\;\;\;\;\;\;\;\;\;\;e^{i\alpha\sqrt{(E^{\prime}+i\kappa)^{2} -\Delta^{2}}} \notag \\
    & + (-2\pi i)\int_{\Delta}^{\infty}dE^{\prime}(-i)\sin(\alpha\sqrt{(E^{\prime} + \hbar\omega)^{2} - \Delta^{2}})e^{i\alpha\sqrt{(E^{\prime} + i\kappa)^{2} -\Delta^{2}}}  
\end{align}
    where we split the result into two domains, because the integrands differ in the limit of $\kappa \rightarrow \infty$. The final result, valid for \textbf{frequencies $\hbar\omega \geq 2\Delta$} is:
    \begin{align}
   I(\omega, R, T = 0) = &(-2\pi i)\int_{\Delta - \hbar\omega}^{-\Delta}dE \left(-\tilde{h}(E)\cos(\alpha\epsilon_{2}) - i\sin(\alpha\epsilon_{2})\right)e^{-i\alpha\tilde{\epsilon}_{1}} \notag  \\
    & + (-2\pi i)\int_{-\Delta}^{0}dE \left(-\tilde{h}(E)\cos(\alpha\epsilon_{2}) - i\sin(\alpha\epsilon_{2})\right)e^{-i\alpha\tilde{\epsilon}_{1}} \notag \\
    +& (-2\pi i)\int_{0}^{\Delta}dE \left(\tilde{h}(E)\cos(\alpha\epsilon_{2}) - i\sin(\alpha\epsilon_{2})\right)e^{i\alpha\tilde{\epsilon}_{1}} \notag \\
    &(-2\pi i)\int_{\Delta}^{\infty}dE\left(\tilde{h}(E)\cos(\alpha\epsilon_{2}) -i\sin(\alpha\epsilon_{2})\right)e^{i\alpha\tilde{\epsilon}_{1}} \notag \\
     & + (2\pi i)\int_{\Delta}^{\infty}\left(\tilde{g}(E)\cos(\alpha\epsilon_{1}) + i\sin(\alpha\epsilon_{1})\right)e^{-i\alpha\tilde{\epsilon}_{2}} 
\end{align}
where we have defined: 
\begin{align}
    \tilde{\epsilon}_{1} = \sqrt{(E + i\hbar\kappa)^{2} - \Delta^{2}} \;\;\;\;\;\; \epsilon_{2} = \sqrt{(E + \hbar\omega)^{2} - \Delta^{2}} \;\;\;\;\;\;
    \tilde{h}(E)  = \frac{(E + \hbar\omega)(E + i\hbar\kappa) + \Delta^{2}}{\epsilon_{2}\tilde{\epsilon}_{1}}
\end{align}
Plugging in $\kappa = 0$ reduces each of $\tilde{\epsilon}_{1}, \tilde{\epsilon}_{2}, \tilde{g}(E), \tilde{h}(E)$ to the functions $\epsilon_{1}, \epsilon_{2}$ \& $g(E)$ defined by Mattis \& Bardeen. 
\begin{center}
    \subsection*{Extreme Anomalous Limit}
\end{center}
We set $\alpha =0$, which considerably simplifies the expression for $I(\omega, R, T)$. Then the complex conductivity is given by (using non-dimensional units - $\omega = \frac{\Delta}{\hbar}\tilde{\omega}$ and $\kappa = \frac{\Delta}{\hbar}\tilde{\kappa}$): 
\begin{align}
    \sigma(\omega, \kappa) = &\frac{1}{\tilde{\omega} - i\tilde{\kappa}}\int_{1 - \tilde{\omega}}^{-1}d\tilde{E}\left(-\frac{(\tilde{E} + \tilde{\omega})(\tilde{E} + i\tilde{\kappa}) + 1}{\sqrt{(\tilde{E} + \tilde{\omega})^{2} - 1}\sqrt{(\tilde{E} + i\tilde{\kappa})^{2} - 1}}\right) \notag \\
    &\frac{1}{\tilde{\omega} - i\tilde{\kappa}}\int_{-1}^{0}d\tilde{E}\left(-\frac{(\tilde{E} + \tilde{\omega})(\tilde{E} + i\tilde{\kappa}) + 1}{\sqrt{(\tilde{E} + \tilde{\omega})^{2} - 1}\sqrt{(\tilde{E} + i\tilde{\kappa})^{2} - 1}}\right) \notag \\
    & + \frac{1}{\tilde{\omega} - i\tilde{\kappa}}\int_{0}^{1}d\tilde{E}\left(\frac{(\tilde{E} + \tilde{\omega})(\tilde{E} + i\tilde{\kappa}) + 1}{\sqrt{(\tilde{E} + \tilde{\omega})^{2} - 1}\sqrt{(\tilde{E} + i\tilde{\kappa})^{2} - 1}}\right)
\end{align}
where we have split up the integration into three different domains as each of the integrands differ in the limit of $\kappa \rightarrow 0^{+}$. Numerical integration showed that the contribution of the last two integrals in $(S79)$ is negligible, hence we have dropped them from further analysis. \\
\subsection*{Conductivity in terms of Elliptic Integrals}
\textbf{Analytical Continuation for $\sigma_{1}$}:
We work with each of the integrals defined above $(S81)$ and apply appropriate transformations to recast the integrals in terms of Incomplete Elliptic Integrals, defined as:
\begin{align}
   \int_{0}^{e^{i\theta}}dx\frac{1}{\sqrt{x^{2}- 1}\sqrt{k^{2}x^{2}-1 }} &= K(e^{i\theta};k) \\
\int_{0}^{e^{i\theta}}dz\frac{\sqrt{1-k^{2}x^2}}{\sqrt{1-x^{2}}} &= E(e^{i\theta}; k)  
\end{align}
where $K(k)$ and $E(k)$ are the complete elliptic integrals of the first and second kind, respectively. In the limit of $\kappa \rightarrow 0^{+}$, the first integral in $(27)$ gives the contribution to $\sigma_{1}(\omega)$. Consider the following integral:
\begin{align}
  \int_{1-\omega}^{-1}dz\frac{(z+\omega)(z + i\kappa) + 1}{\sqrt{(z+\omega)^{2}-1}\sqrt{(z + i\kappa)^{2}-1}}
\end{align}
Apply the transformation: $z \rightarrow t + \lambda$, where $\lambda = \frac{-1}{2}(\omega + i\kappa)$. We also factorize the numerator $(z + \omega)(z + i\kappa) + 1 = z^{2} + i\kappa z + \omega z + i\omega \kappa + 1$ in terms of it's roots:
\begin{align*}
   r_{\pm} = \frac{-(\omega + i\kappa) \pm \sqrt{(\omega - i\kappa)^{2}-4}}{2}
\end{align*}
Applying the transformation, and changing the limits from $1 - \frac{1}{2}(\omega - i\kappa) = \left(1 - \frac{\omega}{2}\right) + \frac{i\kappa}{2}$ to $-1 + \frac{1}{2}(\omega + i\kappa) = \left(\frac{\omega}{2} - 1\right) + \frac{i\kappa}{2}$ gives:
\begin{align*}
    \int_{\left(1 - \frac{\omega}{2}\right) + \frac{i\kappa}{2}}^{\left(\frac{\omega}{2} - 1\right) + \frac{i\kappa}{2}}\frac{(t + \lambda - r_{+})(t + \lambda - r_{-})}{\sqrt{(t + \omega + \lambda)^{2} - 1}\sqrt{(t + \lambda + i\kappa)^{2} - 1}}dt
\end{align*}
The numerator and denominator can be simplified further. 
\begin{align*}
    \lambda - r_{+} &= - \sqrt{\left(\frac{\omega - i\kappa}{2}\right)^{2} - 1} \;\;\;\;
    \lambda - r_{-} = \sqrt{\left(\frac{\omega - i\kappa}{2}\right)^{2} - 1} \\
    \lambda + \omega &= \frac{1}{2}(\omega - i\kappa) \;\;\;\;\;\;\;\;\;\;\;\;\;\;\;\;
    \lambda + i\kappa = -\frac{1}{2}(\omega - i\kappa)
\end{align*}
This gives us the following integral:
\begin{align*}
     \int_{\left(1 - \frac{\omega}{2}\right) + \frac{i\kappa}{2}}^{\left(\frac{\omega}{2} - 1\right) + \frac{i\kappa}{2}} dt \frac{\left(t  - \sqrt{\left(\frac{\omega - i\kappa}{2}\right)^{2} - 1}\right)\left(t + \sqrt{\left(\frac{\omega - i\kappa}{2}\right)^{2} - 1}\right)}{\sqrt{(t + \frac{1}{2}(\omega - i\kappa))^{2} -1 }\sqrt{(t - \frac{1}{2}(\omega - i\kappa))^{2} - 1}}
\end{align*}
To simply further, we apply the transformation $
    t = \mu v = \left(\sqrt{\left(\frac{\omega - i\kappa}{2}\right)^{2} - 1}\right) v $
to obtain the following integral:
\begin{align*}
    \int_{\frac{1}{\mu}\left(1 - \frac{\omega}{2}\right) + \frac{i\kappa}{2\mu}}^{\frac{1}{\mu}\left(\frac{\omega}{2} - 1\right) + \frac{i\kappa}{2\mu}} dv \mu\frac{(v^{2}-1)}{\sqrt{v^{2} - \frac{1}{\mu^{2}}\left(\frac{1}{2}(\omega - i\kappa) -1\right)^{2}}\sqrt{v^{2} -\frac{1}{\mu^{2}}\left(\frac{1}{2}(\omega - i\kappa) + 1\right)^{2}}}
\end{align*}
We now define the parameter $k$ as :
\begin{align}
    k = \frac{\left(\frac{\omega - i\kappa}{2} -1\right)^{2}}{\left(\frac{\omega - i\kappa}{2} - 1 \right)\left(\frac{\omega - i\kappa}{2} + 1\right)} = \frac{\frac{\omega - i\kappa}{2} - 1}{\frac{\omega - i\kappa}{2} + 1}
\end{align}
Simplifications are in order. We note that:
\begin{align*}
    \frac{1}{\mu^{2}}\left(\frac{1}{2}(\omega - i\kappa) -1\right)^{2} &= \frac{\left(\frac{\omega - i\kappa}{2} -1\right)^{2}}{\left(\frac{\omega - i\kappa}{2} - 1 \right)\left(\frac{\omega - i\kappa}{2} + 1\right)} = \frac{\frac{\omega - i\kappa}{2} - 1}{\frac{\omega - i\kappa}{2} + 1}  = k\\
     \frac{1}{\mu^{2}}\left(\frac{1}{2}(\omega - i\kappa) + 1\right)^{2} &= \frac{\left(\frac{\omega - i\kappa}{2}  + 1\right)^{2}}{\left(\frac{\omega - i\kappa}{2} - 1 \right)\left(\frac{\omega - i\kappa}{2} + 1\right)} = \frac{\frac{\omega - i\kappa}{2} + 1}{\frac{\omega - i\kappa}{2} - 1} = \frac{1}{k}
\end{align*}
Then we undertake the final transformation of variables: 
\begin{align}
     v = \sqrt{k}x
\end{align}
The limits of the integration, after the above transformation are:
\begin{align*}
  &\text{Upper Limit:}\;\;\;\;  \frac{\frac{\omega+ i\kappa}{2} -1 }{\sqrt{k}\mu} = \frac{\frac{\omega + i\kappa}{2} -1}{\frac{\omega - i\kappa}{2} -1} = e^{i\theta}
  &\text{Lower Limit}\;\;\;\; \frac{-\frac{\omega- i\kappa}{2}  + 1 }{\sqrt{k}\mu} = -1
\end{align*}
The integral becomes:
\begin{align}
    \mu\sqrt{k}\int_{-1}^{e^{i\theta}}dx\frac{k x^{2}-1}{\sqrt{x^{2}-1}\sqrt{k^{2}x^{2}-1 }} &= \mu\sqrt{k}\int_{0}^{e^{i\theta}}dx\frac{k x^{2}-1}{\sqrt{x^{2}-1}\sqrt{k^{2}x^{2}-1 }} \notag \\ &\;\;\;\; + \mu\sqrt{k}\int_{0}^{1}dx\frac{k x^{2}-1}{\sqrt{x^{2}-1}\sqrt{k^{2}x^{2}-1 }}
\end{align}
where we have used the fact that the integrand is even. Then, we proceed ahead by using Eqs. $(82)$ \& $(83)$ and:
\begin{align*}
    \int_{0}^{1}dx\frac{x^{2}}{\sqrt{x^{2}-1}\sqrt{k^{2}x^{2}-1 }} &= \frac{K(e^{i\theta};k) - E(e^{i\theta};k)}{k^{2}}
\end{align*}
where $K(e^{i\theta};k)$ and $E(e^{i\theta};k)$ are the incomplete elliptic integrals of the first kind and second kind, respectively. Therefore, in terms of elliptic functions,the contribution of the integral in $(81)$ to the conductivity $\tilde{\sigma}(\omega,\kappa)$ is:
\begin{align}
    \boxed{\tilde{\sigma}_{1}(\omega,\kappa) = \left(\frac{1}{2} + \frac{1}{\omega - i\kappa}\right)\left(E(e^{i\theta};k) + E(k)\right) - \frac{2}{\omega - i\kappa}\left(K(e^{i\theta}; k) + K(k)\right)}
\end{align}
\textbf{Analytical Continuation for $\sigma_{2}$:} The integrals contributing to the imaginary conductivity are:
\begin{align*}
    I_{1} &= \int_{-1}^{0}dz \left(-\frac{(z+\omega)(z + i\kappa) + 1}{\sqrt{(z+\omega)^{2}-1}\sqrt{(z + i\kappa)^{2}-1}}\right) \\
    I_{2} &= \int_{0}^{1}dz \left(\frac{(z+\omega)(z + i\kappa) + 1}{\sqrt{(z+\omega)^{2}-1}\sqrt{(z + i\kappa)^{2}-1}}\right)
\end{align*}
Under the transformation $z \rightarrow t + \lambda$, the two integrals can be combined into one, which gives:
\begin{align*}
    I &= \int_{-1-\lambda}^{1 -  \lambda}dt \frac{\left(t  - \sqrt{\left(\frac{\omega - i\kappa}{2}\right)^{2} - 1}\right)\left(t + \sqrt{\left(\frac{\omega - i\kappa}{2}\right)^{2} - 1}\right)}{\sqrt{t^{2} - \left(\frac{1}{2}(\omega - i\kappa) -1\right)^{2}}\sqrt{t^{2} -\left(\frac{1}{2}(\omega - i\kappa) + 1\right)^{2}}}
\end{align*}
We now undertake the same transformations as before. The transformation, $t = \mu v$ gives:
\begin{align*}
    I &= \int_{\frac{-1-\lambda}{\mu}}^{\frac{1-\lambda}{\mu}}\mu dv \frac{v^{2}-1}{\sqrt{v^{2} - \frac{1}{k}}\sqrt{v^{2} - k}}
\end{align*}
Then, the transformation: $v = \sqrt{k}x$ gives:
\begin{align*}
    I &= \int_{\frac{-1-\lambda}{\mu\sqrt{k}}}^{\frac{1-\lambda}{\mu\sqrt{k}}} dx \mu \sqrt{k}\frac{k x^{2}-1}{\sqrt{k^{2}x^{2}-1}\sqrt{x^{2}-1}}
\end{align*}
with the limits:
\begin{align*}
    \frac{1-\lambda}{\mu \sqrt{k}} &= \frac{\frac{\omega + i\kappa}{2} + 1}{\frac{\omega - i\kappa}{2}-1} = k_{1}\\
    \frac{-1-\lambda}{\mu \sqrt{k}} &= \frac{\frac{\omega + i\kappa}{2}-1}{\frac{\omega - i\kappa}{2}-1} = e^{i\phi}
\end{align*}
where have that $|k_{1}| > 1$. Consider the following integral:
\begin{align*}
    \int_{e^{i\phi}}^{k_{1}}dx\frac{1}{\sqrt{k^{2}x^{2}-1}\sqrt{x^{2}-1}} &= \int_{0}^{k_{1}}dx \frac{1}{\sqrt{k^{2}x^{2}-1}\sqrt{x^{2}-1}} - \int_{0}^{e^{i\phi}}dx\frac{1}{\sqrt{k^{2}x^{2}-1}\sqrt{x^{2}-1}} \\
    &= K(k)  - K(e^{i\phi};k) -\textcolor{red}{i}K(k^{\prime}) + \textcolor{red}{i}K\left(\frac{\sqrt{1-k^{2}k_{1}^{2}}}{k^{\prime}};k^{\prime}\right)
\end{align*}
where $K(k_{1};k)$ represents the incomplete elliptic integral of the first kind.\\\\ 
The contribution of the above integral to $I$ is:
\begin{align}
    (-i)\left(\frac{\omega - i\kappa}{2} - 1\right)\left\{iK(k)  - iK(e^{i\phi};k)+K(k^{\prime}) - K\left(\frac{\sqrt{1-k^{2}k_{1}^{2}}}{k^{\prime}};k^{\prime}\right)\right\}
\end{align}
Consider the remaining integrals:
\begin{align*}
&\int_{e^{i\phi}}^{k_{1}} dx \frac{kx^{2}}{\sqrt{k^{2}x^{2}-1}\sqrt{x^{2}-1}} = \int_{0}^{k_{1}}dx \frac{kx^{2}}{\sqrt{k^{2}x^{2}-1}\sqrt{x^{2}-1}} - \int_{0}^{e^{i\phi}}dx\frac{kx^{2}}{\sqrt{k^{2}x^{2}-1}\sqrt{x^{2}-1}}   
\end{align*}
We note that the integrand above can be split as:
\begin{align*}
    \frac{kx^{2}}{\sqrt{1-k^{2}x^{2}}\sqrt{1-x^{2}}} = \frac{1}{k}\left(\frac{1}{\sqrt{1-k^{2}x^{2}}\sqrt{1-x^{2}}} - \frac{\sqrt{1-k^{2}x^{2}}}{\sqrt{1-x^{2}}}\right)
\end{align*}
which leads us to consider the following integrals:
\begin{align*}
    \frac{1}{k}\int_{e^{i\phi}}^{k_{1}}dx\frac{1}{\sqrt{1-k^{2}x^{2}}\sqrt{1-x^{2}}} = \frac{K(k)}{k} -i\frac{K(k^{\prime})}{k} + \frac{i}{k}K\left(\frac{\sqrt{1-k^{2}k_{1}^{2}}}{{k^{\prime}}};k^{\prime}\right) - \frac{K(e^{i\phi};k)}{k}
\end{align*}
\begin{align*}
    \frac{1}{k}\int_{e^{i\phi}}^{k_{1}}dx\frac{\sqrt{1-k^{2}x^{2}}}{\sqrt{1-x^{2}}} &= -i\frac{K(k^{\prime})}{k} + i\frac{E(k^{\prime})}{k} +\frac{i}{k}K\left(\frac{\sqrt{1-k_{1}^{2}k^{2}}}{k^{\prime}};k^{\prime}\right) - \frac{i}{k}E\left(\frac{\sqrt{1-k_{1}^{2}k^{2}}}{k^{\prime}};k^{\prime}\right)  \\ & \;\;\;\;\;\;+ \frac{E(k)}{k} - \frac{E(e^{i\phi};k)}{k}
\end{align*}
We can simplify the expressions by subtracting the above two terms, giving us: 
\begin{align*}
    &\frac{1}{k}\int_{e^{i\phi}}^{k_{1}}dx\frac{1}{\sqrt{1-k^{2}x^{2}}\sqrt{1-x^{2}}} -  \frac{1}{k}\int_{e^{i\phi}}^{k_{1}}dx\frac{\sqrt{1-k^{2}x^{2}}}{\sqrt{1-x^{2}}} = -\frac{i}{k}E(k^{\prime}) + \frac{i}{k}E\left(\frac{\sqrt{1-k_{1}^{2}k^{2}}}{k^{\prime}};k^{\prime}\right) \\
    &\;\;\;\;\;\; + \frac{K(k) - K(e^{i\phi};k)}{k} + \frac{E(k) - E(e^{i\phi};k)}{k}  
\end{align*}
The contribution to $I$ is:
\begin{align}
    (-i)\left(\frac{\omega - i\kappa}{2}+ 1\right)\left\{E(k^{\prime}) - E\left(\frac{\sqrt{1-k_{1}^{2}k^{2}}}{k^{\prime}};k^{\prime}\right) + iK(k) - iK(e^{i\phi};k) + iE(k) - iE(e^{i\phi};k)\right\}
\end{align}
Therefore, the analytically continued form for $(-i)\sigma_{2} = \frac{I}{\omega - i\kappa}$ is:
\begin{align}
    &\frac{1}{2}\left\{\left(\frac{2}{\omega - i\kappa} + 1\right)E(k^{\prime}) + \left(\frac{2}{\omega - i\kappa} - 1\right)K(k^{\prime})\right\} + \frac{2i}{\omega - i\kappa}\left\{K(k) - K(e^{i\phi};k)\right\} \notag \\
    & + \frac{i}{2}\left(\frac{2}{\omega-i\kappa} + 1\right)\left\{E(k) - E(e^{i\phi};k)\right\} - \frac{1}{2}\left(\frac{2}{\omega-i\kappa} + 1\right)E\left(\frac{\sqrt{1-k_{1}^{2}k^{2}}}{k^{\prime}};k^{\prime}\right) \notag \\
    &  - \frac{1}{2}\left(\frac{2}{\omega-i\kappa} - 1\right)K\left(\frac{\sqrt{1-k_{1}^{2}k^{2}}}{k^{\prime}};k^{\prime}\right)
\end{align}
\bibliographystyle{apsrev4-2}
\bibliography{suppref}